\documentclass[10pt,aps,prx,showpacs,superscriptaddress,twocolumn,citeautoscript]{revtex4-1}

\usepackage{graphicx}  % needed for figures
\usepackage{dcolumn}   % needed for some tables
\usepackage{bm}        % for math 
\usepackage{amssymb}  
\usepackage{amsmath,braket}
\usepackage{hyperref}
\usepackage[usenames]{color}

\newcommand{\beq}{\begin{equation}}
\newcommand{\eeq}{\end{equation}}

\newcommand{\fr}{\bm{r}}
\newcommand{\fx}{\bm{\hat{x}}}

\newcommand{\fK}{\bm{k}}
\newcommand{\fu}{\bm{u}}

\newcommand{\fn}{\bm{n}}
\newcommand{\fE}{\bm{E}}

\newcommand{\fH}{\bm{H}}
\newcommand{\fB}{\bm{B}}
\newcommand{\fD}{\bm{D}}
\newcommand{\fJ}{\bm{J}}
\newcommand{\fP}{\bm{P}}

\def\bean{\begin{eqnarray*}}
\def\eean{\end{eqnarray*}}

\newcommand{\eps}{\varepsilon}
\newcommand{\subeqs}[1]{\begin{align} #1 \end{align}}
\begin{document}
\title{Generalized Elastodynamic Model for Nanophotonics }
\author{J.V. Alvarez}
\affiliation{Departamento de F\'isica de la Materia Condensada, Universidad Aut\'onoma de Madrid, Madrid 28049, Spain
and Condensed Matter Physics Center (IFIMAC) and Instituto Nicol\'as Cabrera (INC)}
\author{Bahram Djafari-Rouhani}
\affiliation{Institut d'Electronique, de Microl\'ectronique et de Nanotechnologie, UMR CNRS 8520, Universit\'e de Lille 
1, 59655 Villeneuve d'Ascq, France}

\author{Dani Torrent}
\email{dtorrent@uji.es}
\affiliation{GROC, UJI, Institut de Noves Tecnologies de la Imatge (INIT), Universitat Jaume I, 12071, Castell\'o, (Spain)}

\date{\today}
%%%%%%%%%%%%%%%%%%%%%%%%%%%%%%%%%%%%%%%%%%%%%%%%%%%%%%%%%%%%%%%%%%%%%%%%%%%%%%%%%%%%%%%%%%%

%%%%%%%%%%%%%%%%%%%%%%%%%%%%%%%%%%%%%%%%%%%%%%%%%%%%%%%%%%%%%%%%%%%%%%%%%%%%%%%%%%%%%%%%%%%
\begin{abstract}
A self-consistent theory for the classical description of the interaction of light and matter at the nano-scale is presented, which takes into account spatial dispersion. Up to now, the Maxwell equations in nanostructured materials with spatial dispersion have been solved by the introduction of the so-called Additional Boundary Conditions (ABC) which, however, lack generality and uniqueness. In this paper, we derive an approach where non-local effects are studied in a precise and uniquely defined way, thus allowing the treatment of all solid-solid interfaces (among metals, semiconductors or insulators), as well as solid-vacuum interfaces in the same framework. The theory is based on the derivation of a potential energy for an ensemble of electrons in a given poptential, where the deformation of the ensemble is treated as in a solid, including both shear and compressional deformations, instead of a fluid described only by a bulk compressibilty like in the hydrodynamical approach. The derived classical equation of motion for the ensemble describes the deformation vector and the corresponding polarization vector as an elastodynamic field, including viscous forces, from which a generalized non-local constitutive equation for the dielectric constant is derived. The required boundary conditions are identical to that of elastodynamics and they emerge in a natural way, without any physical hypothesis outside the current description, as it is commonly required in other non-local approaches. Interestingly, this description does not require the discontinuity of any component of the electric, magnetic or polarization fields and, consequently, no bounded currents or charges are present at the interface, which is a more suitable description from the microscopic point of view. It is shown that the method converges to the local boundary conditions in the low spatial dispersion limit for insulators and conductors, quantified by means of a parameter defined as the « characteristic length ». A brief discussion about the inclusion of the spill-out of electrons across surfaces is discussed. Finally, the scattering by a plane, a cylinder and a sphere is studied and numerical examples of the behaviour of the different fields at the interfaces are presented, showing the limiting situations in which the local limit is recovered, reinforcing the self-consistency of this description.
\end{abstract}
%%%%%%%%%%%%%%%%%%%%%%%%%%%%%%%%%%%%%%%%%%%%%%%%%%%%%%%%%%%%%%%%%%%%%%%%%%%%%%%%%%%%%%%%%%%
\maketitle
%%%%%%%%%%%%%%%%%%%%%%%%%%%%%%%%%%%%%%%%%%%%%%%%%%%%%%%%%%%%%%%%%%%%%%%%%%%%%%%%%%%%%%%%%%%
\section{Introduction}
%%%%%%%%%%%%%%%%%%%%%%%%%%%%%%%%%%%%%%%%%%%%%%%%%%%%%%%%%%%%%%%%%%%%%%%%%%%%%%%%%%%%%%%%%%%
%The evolution of the electromagnetic field in matter from a classical perspective is described by means of Maxwell's equations, but these have to be complemented by constitutive equations relating the induced current with the electromagnetic field. The simpler form of the constitutive equations assumes a local and linear relationship between the induced current and the electric and magnetic fields, from which we define the well-known dielectric permittivity and magnetic permeabilities\cite{jackson1999classical}.

% \textcolor{blue}{different from...(You propose 2 options CO and SC, but which one doesn't fit?) } 
The interaction of light and matter at the nanoscale has been a topic of intense research in recent years\cite{kelly2003optical,amendola2017surface}, due to the extraordinary advances in the manipulation capabilities of materials at the nanoscale\cite{burda2005chemistry,matricardi2018gold,espinha2018hydroxypropyl}. The range of applications of this science is extremely broad and continues growing up\cite{raza2015nonlocal}, therefore theoretical and numerical tools for their accurate description are of primary interest\cite{aizpurua2003optical,myroshnychenko2008modelling,baumberg2019extreme}.

The nanoscale is a complex size limit for the study of the interaction of light and matter, since typical structures are big enough to consider the problem from the classical point of view although some quantum effects can be observable. However, it is still possible to use a full classical description, as long as we find a proper constitutive equation relating the electric and magnetic fields with the induced currents and charges, which can take into account quantum corrections\cite{ciraci2013hydrodynamic}.

The simpler form of the constitutive equations used in electrodynamics are linear and local in both space and time, and they define the dielectric permittivity, magnetic permeability and electrical conductivity\cite{jackson1999classical}. Although nonlocality in time is commonly assumed, resulting in frequency-dependent constitutive parameters, spatial non-locality is in general left behind to more refined models of matter, since they become important only at the nanoscale\cite{adler1962quantum,agranovich2013crystal,agranovich2006spatial,garcia2008nonlocal}. It has to be pointed out that in the domain of metamaterials spatial dispersion has been a topic of intense research as well, since the distance between the ``meta-atoms'' is only one order of magnitude smaller than the operating wavelength of the field\cite{belov2003strong,menzel2010validity,alu2011restoring,chipouline2012basics,mnasri2018beyond,mnasri2019retrieving,mnasri2019homogenization}.

Despite the great success of the spatially local description at the macro or even micro scales, the theory fails in the accurate description of nanomaterials, since spatial dispersion becomes more relevant and it has to be included in the constitutive parameters\cite{rojas1988nonlocal,pack2001failure,silveirinha2006additional,mcmahon2010nonlocal,maslovski2010generalized,david2011spatial,moreau2013impact,gomez2016hierarchical}.

When the constitutive equations become spatially non-local additional modes emerge in the solution of the wave equation, and the boundary conditions derived within the framework of Maxwell's equations are insufficient to match all the excited fields at an interface between two materials. This problem has derived in a countless number works proposing the so-called ``additional boundary conditions'', which are different depending on the different response models\cite{pekar1958theory,melnyk1970theory,halevi1984generalised,gomez2016hierarchical,yang2019general}, what in turn means that spatial dispersion is not generally treated in the same way in insulators, semiconductors or metals. Despite the fact that these additional boundary conditions are deduced with more or less reasonable physical assumptions, they are not derived within the framework of Maxwell's equations complemented with the constitutive equations, what means that the description is not ``closed'' and it can result in an inaccurate description of the electrodynamic problem, as can be seen for the fact that the problem still remains unsolved\cite{henneberger1998additional,luo2013surface,kupresak2018comparison}.

%There is an additional drawback in the description of electrodynamics in matter that has been less taken into account in the literature, as is the discontinuity of the induced current at an interface, which results in discontinuities of the electromagnetic field and the existence of bounded charges and currents. It is commonly assumed that these discontinuities and surface fields are idealizations properly justified at the macroscale, and they are overcome with the definitions of some macroscopic fields which are continuous at the interfaces. However, from the microscopical point of view a description in which the induced current and charges be continuous functions would be more satisfactory, since the carriers of these fields are entities (electrons, holes and other excitons) which are described by a wavefunction which is continuous through interfaces.

In this work we will show that, by means of an elastodynamic model of the induced current, we can find a self-consistent description of the light-matter interaction which accounts for spatial dispersion. We will show that when an ensemble of electrons is deformed its potential energy is identical to that of an elastic body, so that dynamically it moves as a continuous elastodynamic field\cite{royer1999elastic}. In this context, we can develop a non-local theory for electrodynamics, where the required boundary conditions arise in a  natural way. Moreover, we will define the local limit by means of a ``characteristic length'' parameter in which we recover the local description, reinforcing therefore the generality of this model.

The paper is organized as follows. After this introduction, in section \ref{sec:preliminary}  a discussion about the problem of non-locality and its possible solutions is presented, then section \ref{sec:QS} presents the notions of quantum pressure and stress, and the elastic energy of the ensemble of electrons is derived. Section \ref{sec:EF} presents the elastodynamic formulation for nanophotonics, with a discussion about boundary conditions. Section \ref{sec:ISO} analyzes the solution of the wave equations in isotropic materials. Section \ref{sec:SO} discusses how this approach can include the spill-out of electron across surfaces and section \ref{sec:HE} compares the present approach with the hydrodyamic description. Section \ref{sec:SC} analyzes the solid-solid and vacuum-solid interfaces for planar, cylindrical and spherical geometries. Finally, section \ref{sec:SM} summarizes the work.
%%%%%%%%%%%%%%%%%%%%%%%%%%%%%%%%%%%%%%%%%%%%%%%%%%%%%%%%%%%%%%%%%%%%%%%%%%%%%%%%%%%%%%%%%%%
%\section{Preliminary discussion: Local and non-local materials}
\section{Local and non-local materials}
%%%%%%%%%%%%%%%%%%%%%%%%%%%%%%%%%%%%%%%%%%%%%%%%%%%%%%%%%%%%%%%%%%%%%%%%%%%%%%%%%%%%%%%%%%%
\label{sec:preliminary}

The evolution of the electromagnetic field inside a material where no sources are present is described by means of Maxwell's equations\cite{jackson1999classical},
\subeqs{
\nabla\times\fE&=-\partial_t\fB, \label{eq:rotE}\\
\nabla\times\fB&=\mu_0\eps_0\partial_t\fE+\mu_0\fJ_i, \label{eq:rotB} \\
\nabla\cdot\fB&=0, \label{eq:divB} \\
\nabla\cdot\fE&=\rho_i/\eps_0, \label{eq:divE}
}
where $\fJ_i$ and $\rho_i$ are the induced current and charge densities in the material, respectively. The continuity equation for the current density is implicit in Maxwell's equations, and it is derived taking the divergence of equation \eqref{eq:rotB} and using equation \eqref{eq:divE}, 
\beq
\label{eq:cont}
\nabla\cdot\fJ_i+\partial_t\rho_i=0.
\eeq

Maxwell's equations with the induced currents are not enough to solve the full electrodynamic problem, and we need a relationship between the induced current $\fJ_i$ and the electromagnetic field. This relationship is complex to obtain, since it implies a many-body problem solved in the framework of classical, semiclassical or quantum theories, however a phenomenological approach is commonly considered in which the classical permittivity, conductivity and permeability are defined. Then, in classical electrodynamics, for non-magnetic and non-chiral materials, the induced current is expressed as
\beq
\label{eq:constitutive}
\fJ_i=\sigma_E \fE+\eps_0\chi_E\partial_t\fE,
\eeq
%
%\subeqs{
%\fJ_F&=\sigma \fE\\
%\fP&=(\eps-\eps_0)\fE\\
%\bm{M}&=(\mu^{-1}-\mu_0)\fB
%}
%
%\beq
%\fJ=(\sigma-i\omega(\eps-\eps_0)\fE+\frac{i}{\omega}\left(\frac{\mu}{\mu_0}-1\right)\left(\fK\fK\cdot\fE-k^2\fE\right)
with $\sigma_E$ and $\chi_E$ being the electric conductivity and susceptibility, respectively. It is common to define the polarization vector $\fP$ as
\beq
\fJ_i=\partial_t\fP,
\eeq
so that, assuming a harmonic time dependence of the fields of the form $\exp(-i\omega t)$, equation \eqref{eq:constitutive} is
\beq
\label{eq:Pconst}
\fP=(\eps_0\chi_E+i\sigma_E/\omega)\fE,
\eeq
which defines the well known complex dielectric constant $\eps$ by means of the dielectric displacement $\fD$, defined as
\beq
\fD=\eps_0\fE+\fP=\eps\fE,
\eeq
so that
\beq
\eps=1+\chi_E+i\sigma_E/(\eps_0\omega).
\eeq

If either $\sigma_E$ or $\chi_E$ are discontinuous at an interface, so it is the current $\fJ_i$ and, by means of equation \eqref{eq:cont}, a surface charge appears. In classical electrodynamics these discontinuities and surface fields are usual, and they do not imply any non-physical description of the fields.

However, there are two points which make equation \eqref{eq:constitutive} unpleasant from a deeper physical insight. First, this local relationship in both time and space implies an instantaneous response of matter (locality in time) and a point-to-point response (locality in space), that is, the induced current in the material is instantaneous and it depends on the field at a given point only. Clearly this is not a true physical situation, since the response of the charges will have some inertial response and will be influenced by the surrounding material. Another point, less discussed in the literature, is that current discontinuities and surface charges are not really possible at the microscopic level, since charges and currents are quantum entities described by wave functions which are in general continuous across the interfaces. Then, while it is true that the classical description of electrodynamics assumes that these discontinuities are just idealizations, we would like to find a description in which these do not occur, so that the theory would allow us to reduce more and more the scale of validity. 

In this work we will show how, surprisingly, a non-local version of equation \eqref{eq:constitutive} is possible which additionally implies the continuity of all the fields involved in the interaction.

\subsection{Non-local constitutive equation}
When elementary models of the light-matter interaction are considered,  the derived constitutive parameters $\chi_E$ and $\sigma_E$ are found to be frequency-dependent, but this dependence has no consequences in the nature of the fields, since equations \eqref{eq:rotE} to \eqref{eq:divE} are usually solved assuming time-harmonic dependence.

However, the constitutive parameters can depend on the wavenumber as well, and this dependence is not equivalent to the frequency-dependence since, usually, we have to work on bounded materials. Then, let us assume for instance that the dielectric constant is both frequency and wavenumber dependent. We will have (for the purposes of this section we will consider only a scalar-like relation between fields, the full vectorial theory is developed later on) 
\beq
D(k,\omega)=\eps(k,\omega)E(k,\omega),
\eeq
while we can continue working on the frequency-domain, if our material is bounded we need to work on the space domain, and Fourier transform the above equation, thus we have
\beq
D(x,\omega)=\int_{-\infty}^\infty \eps(x-x',\omega)E(x',\omega)dx',
\eeq
where we have used the convolution theorem. The above equation clarifies the term non-local for wavenumber-dependent constitutive paramaters: the dielectric displacement at point $x$ depends linearly on the electric field at point $x'$, distant from $x$ a length $d=x-x'$. The dielectric function   $\eps(d,\omega)$, as a non-local response function, weights the contribution of electric fields at different $x'$,  and will decay as  $d$ increases.  In the local limit, this function is proportional to the Dirac delta function $\delta(x-x')$. It is obvious that, if we introduce this relationship of the dielectric displacement with the electric field in equations \eqref{eq:rotE} to \eqref{eq:divE}, even in the time-harmonic regime we obtain a more complicated equation. As the convolution theorem applies only to a spatially invariant medium, for a bounded material the dielectric constant will not be in general a function of $d=x-x'$ and  we get the most general linear relation between two scalar fields.
\beq
\label{eq:DEgeneral}
D(x,\omega)=\int_{-\infty}^\infty \eps(x,x',\omega)E(x',\omega)dx',
\eeq
which complicates even more the problem. In next subsection we will discuss how this problem could be solved in Fourier space by means of the so called ``additional boundary conditions''.

\subsection{The need of additional boundary conditions}
The usual method to solve the electrodynamic equations in bounded materials is by means of the application of boundary conditions, roughly speaking: knowing the solution of the fields in two materials, we apply boundary conditions at the interface between them and we find the solution of the problem. Maxwell's equations \eqref{eq:rotE} to \eqref{eq:divE} provide us of the required number of boundary conditions, once the constitutive equation for the induced current has been found, but only if this constitutive equation is local in space, otherwise we require of additional boundary conditions, and these additional boundary conditions has been (and continues being) a topic of a great discussion in the literature. The origin of this need is found in the number of solutions that a wave-number dependent dielectric constant provides, as will be explained below.

Let us assume the problem of reflection and transmission at a flat interface. If the dielectric constant is frequency-dependent only, the dispersion relation in the material is typically of the form
\beq
k^2=\mu_0\eps(\omega)\omega^2,
\eeq
which means that, for a given frequency, we have two waves propagating through oposite directions (at normal incidence and in this scalar version, the vectorial case is more complex, as will be explained later). For a classical reflection and transmission problem we will have therefore the incident wave, the reflected wave propagating backwards and the transmitted field at the other material. We will need to determine the amplitude of the reflected and transmitted waves and Maxwell equations provide us indeed two boundary conditions: the problem is perfectly defined. 

However, if the dielectric constant is also wavenumber-dependent, for a given frequency our dispersion relation is
\beq
k^2=\mu_0\eps(\omega,k)\omega^2,
\eeq
which, due to the dependence on $k$ of $\eps$, can give more than two modes propagating in opposite directions. Then, a given incident field will excite, in the simpler of the situations, two reflected and two transmitted modes (assuming both materials non-local).

Maxwell equations provides only two boundary conditions, but we have four modes to determine, so that the problem is not well defined. We need the so-called ``additional boundary conditions''  to completely solve the problem.

The amount of works about these additional boundary conditions  is huge and actually experts have developed the acronym ABC to refer to specific examples or generally, to the very issue presented above.  There is no consensus about which ones are the correct ones. It has been assumed then that the right boundary conditions depend on the micro-structure of the materials and the interfaces, so that there is not a unique solution for the problem of additional boundary conditions at the macroscopic level. However, we believe there is a well-defined  macroscopic solution to this problem based on a right definition of the constitutive parameters, whose specific values of course depend on the micro-structure of the material but whose macroscopic behaviour can be universally defined, as we do in the electrodynamics of local materials, where obviously we don't apply the same boundary conditions for metals and for lossless dielectrics, but it is due to the fact that their constitutive parameters have different values. Next subsection will present the approach we will follow in this work to overcome the problem of the additional boundary conditions.

\subsection{Solution from additional field equations}
The additional boundary conditions have been discussed in uncountable ways, and these are suggested by more or less acceptable physical arguments or microscopical models. Our objective is to find a set of macroscopic equations from which obtain the required boundary conditions, with these equations being functions of a set of (local) constitutive parameters whose numerical values define the different type of materials. It is obvious that the additional boundary conditions cannot be obtained from Maxwell equations, and physical arguments can be enough for simple situations (vacuum-solid interface at normal incidence), but more general situations require of a more rigorous solution.

The motivation to search for the field equation of the induced current comes from equation \eqref{eq:DEgeneral}, which is equivalent to a relationship between the polarization $\fP$ and the electric field $\fE$ of the form
\beq
\fP(\fr,\omega)=\eps_0\iiint\chi_E(\fr,\fr',\omega)\fE(\fr',\omega)dV.
\eeq

The above equation suggests indeed that the function $\chi_E(\fr,\fr',\omega)$ is the Green's function of differential equation of the form
\beq
\mathcal{L}\fP=\eps_0\fE
\eeq
and
\beq
\mathcal{L}\chi_E(\fr,\fr',\omega)=\delta(\fr-\fr'),
\eeq
with $\mathcal{L}$ being a partial differential operator. If we are able to find the operator $\mathcal{L}$ we will be able to find the right boundary conditions in a similar way as we find them from Maxwell equations. 

%In principle it could be thought that the operator $\mathcal{L}$ might depend on the polarization carrier, since at the microscopic level the polarization can be due to free or bounded electrons, holes, phonons, etc. However, we will discuss in next section that the form of the  $\mathcal{L}$ operator is similar in the main polarization carriers, at least in the first order approximation, and we will focus our discussion in the free carriers (electrons and holes).

It has to be pointed out that the objective of this work is not to model nor discuss the consequences of the frequency-dependence of the complex dielectric constant, our objective is to discuss the consequences of the wave-number dependence, which will enter into the model through spatial derivatives in $\mathcal{L}$, and how this affects boundary conditions. Then, it has to be assumed that all the constants appearing in the model equations could present a more or less complex frequency dependence, even more if our material is artificially nano-structured, but we will not take care of this dependence.
%%%%%%%%%%%%%%%%%%%%%%%%%%%%%%%%%%%%%%%%%%%%%%%%%%%%%%%%%%%%%%%%%%%%%%%%%%%%%%%%%%%%
\section{Quantum Stress-Strain Relation}
\label{sec:QS}

The most elementary microscopic model of the dielectric constant assumes that the electron is bounded to the atom by means of a spring-like restoring force\cite{jackson1999classical}, so that when the electric field interacts with it the equation of motion is
\beq
\label{eq:particle}
m_e \ddot{\fr}+\gamma \dot{\fr}+\kappa \fr=-e\fE,
\eeq
from which we can solve for the contribution to polarizability of one single electron as $\bm{p}=-e\fr$. If the material is a conductor the restoring force $-\kappa \fr$ is set to zero and we recover Drude's model for the free electron. While these models are quite elementary, they allow to explain some aspects of light-matter interaction, and in both models we recover a local and complex dielectric function. 

\subsection{Quantum pressure and quantum stress}
Sommerfeld's model applies quantum statistics to the physics of the free electron in the solid, which is described as an ensemble of non-interacting spin 1/2 particles. It is found that an ensemble of $N$ particles in a box of volume $\Omega$ has a total energy $E$ given by
\beq
E=\frac{3}{5}NE_F,
\eeq
where $E_F$ is Fermi's energy and is the maximum energy level occupied by the electrons. This energy is proportional to $n^{2/3}$, with $n=N/\Omega $, so that we can obtain the so-called quantum pressure $\mathcal{P}$ of the gas from the thermodynamical definition of pressure,
\beq
\label{eq:PThermo}
\mathcal{P}=-\left(\frac{\partial E}{\partial \Omega}\right)_N=\frac{2}{3}\frac{E}{\Omega},
\eeq
and the compressibility or bulk modulus
\beq
\label{eq:Bulk}
B=-\Omega\frac{\partial P}{\partial \Omega}=\frac{5}{3}\mathcal{P}.
\eeq
The electron gas is then described as a fluid material and the linearized equation of motion is that of the acoustic field subject to a body pressure due to the external electric field. This is the so-called hydrodynamic model for plasmonics\cite{ciraci2013hydrodynamic} and, although widely used, we will show here that a more accurate description is needed, since the electron gas is found to have not only a bulk but also a shear modulus, so that it is better described as a solid material.

Other models also include in this equation Ohmnic losses or diffusion\cite{mortensen2014generalized}, what essentially changes the frequency dependence of the constants involved in the model. Including other sources of dissipation at the quantum level, inter or intra band transitions for example\cite{fox2002optical}, might also change this frequency response, but for the purpose of this work it will be enough to assume that all the parameters appearing in the models are complex and frequency-dependent, since our objective is not to model the interaction, but to understand the role and nature of the additional spatial derivatives appearing in the equation of motion for the induced current.

The hydrodynamic model assumes that the gas of electrons is a fluid, and that only volumetric changes are possible. Moreover, it assumes as well that the internal restoring force is due to the gradient of a scalar pressure field. We will assume here a more general deformation of the ensemble of electrons to demonstrate that they behave actually as a solid.

Let us assume we have an ensemble of electrons in equilibrium in a periodic potential $V(\fr)$. Let us assume now that some external perturbation (like an electric field) modifies this wave function so that a deformation $\fu(\fr)$ is applied, in such a way that the coordinates are transformed as
\beq
\fr'=\fr+\fu(\fr).
\eeq
If a strain is defined in the usual way
\beq
\epsilon_{ij}=\frac{1}{2}\left(\frac{\partial u_i}{\partial x_j}+\frac{\partial u_j}{\partial x_i}\right),
\eeq
the stress of the deformed system is found as
\beq
\label{eq:sigmaE}
\sigma_{ij}=\frac{\partial \mathcal{E}}{\partial \epsilon_{ij}},
\eeq
with $\mathcal{E}$ being the elastic energy density. It is important to remark that, if rotations are excluded from the dynamics, the strain tensor is symmetric and the following property holds
\beq
\frac{\partial \epsilon_{k\ell}}{\partial \epsilon_{ij}}=\delta_{ik}\delta_{j\ell}+\delta_{i\ell}\delta_{jk}-\delta_{ij}\delta_{k\ell}.
\label{derivativeitself}
\eeq

Let us assume now that the Hamiltonian of the system is given by
\beq
\mathcal{H}=\sum_{\alpha}\left[\frac{\bm{p}_\alpha^2}{2m_\alpha}+V(\fr_\alpha)+\frac{1}{2}\sum_{\beta\neq\alpha}U(r_{\alpha\beta})\right],
\eeq
where $\bm{p}_\alpha=-i\hbar\nabla_\alpha$, $\fr_\alpha$ and $m_\alpha$ are the momentum, position and mass of the $\alpha$ particle, respectively. The term $V(\fr_\alpha)$ stands for the periodic potential of the lattice.  The term  $U(r_{\alpha\beta})$, whith $r_{\alpha\beta}=|\mathbf{r}_\alpha- \mathbf{r}_\beta|$, is a two-body interaction between electrons.

Once the transformation is applied, each of these terms will contribute in a different way to the energy of the system. If a small deformation is assumed, the energy of the deformed system will be a function of the deformation $\fu$ and the strain $\nabla\fu$. The difference between the energy of the deformed and undeformed systems is called the elastic energy, and it will allow us to define the elastic constants of the ensamble and to deduce the equation of motion.

%%%%%%%%%%%%%%%%%%%%%%%%%%%%%%%%%%%%%%%%%%%%%%%%%%%%%%%%%%%%%%%%%%%%%%%%%%%%%%
 \subsection{Kinetic Term and the Shear Modulus of the Free Electron Gas}
The kinetic energy of the deformed system, using the transformation $\partial_i'=\partial_i-\partial_ju_i\partial_j$, is given by
\beq
E_K=\frac{1}{2m_e}\braket{\bm{p}^2-2\frac{\partial u_i}{\partial x_j}p_{ i}p_{ j}+\frac{\partial u_i}{\partial x_j}\frac{\partial u_i}{\partial x_k}p_{ j}p_{ k}}.
\eeq

The first term of the right hand side part of the above equation is the kinetic energy of the undeformed system,
\beq
E_K^0=\braket{\frac{\bm{p}^2}{2m_e}},
\eeq
and the second term will be
\beq
-\frac{1}{m_e}\braket{\frac{\partial u_i}{\partial x_j}p_{ i}p_{ j}}=-\braket{\frac{p_i^2}{m_e}}\frac{\partial u_i}{\partial x_i}=-\frac{2}{3}E_K^0\nabla\cdot\fu,
\eeq
where we have used the equipartition energy theorem, 
\beq
\braket{p^2_i/2m}=1/3\braket{\bm{p}^2/2m}.
\eeq
Finally, the last term is

\beq
\braket{\frac{1}{2m_e}\frac{\partial u_i}{\partial x_j}\frac{\partial u_i}{\partial x_k}p_{ j}p_{ k}}=\frac{1}{3}E_K^0\sum_{i,j}\left(\frac{\partial u_i}{\partial x_j}\right)^2,
\eeq
so that the contribution to the elastic energy of the kinetic part is given by
\beq
E=-\frac{2}{3}E_K^0\nabla\cdot\fu+\frac{1}{3}E_K^0\sum_{i,j}\left(\frac{\partial u_i}{\partial x_j}\right)^2.
\eeq
If rotations are ignored, the energy density can be expressed in terms of the strain as
\beq
\mathcal{E}=\frac{E_K^0}{\Omega}\left(-\frac{2}{3}\epsilon_{ii}+\frac{1}{3}\epsilon_{ij}\epsilon_{ij}\right).
\eeq

The volume $\Omega$ is also a function of the strain, and is
\beq
\Omega=\Omega_0(1+\epsilon_{jj}),
\eeq
with $\Omega_0$ being the volume of the undeformed system. The energy density is therefore given by
\beq
\mathcal{E}\approx\frac{E_K^0}{\Omega_0}\left(-\frac{2}{3}\epsilon_{ii}+\frac{2}{3}\epsilon_{ii}\epsilon_{jj}+\frac{1}{3}\epsilon_{ij}\epsilon_{ij}\right)  
\eeq
from which we can obtain the stress of the system using equation \eqref{eq:sigmaE}
\beq
\sigma_{ij}=\frac{\partial \mathcal{E}}{\partial \epsilon_{ij}}=\sigma_{ij}^0+C_{ijk\ell}\epsilon_{k\ell},
\eeq
where
\beq
\label{eq:sigmaP0}
\sigma_{ij}^0=-\frac{2}{3}\frac{E_K^0}{\Omega_0}\delta_{ij},
\eeq
and
\beq
\label{eq:Cfree}
C_{ijk\ell}=\frac{2}{3}\frac{E_K^0}{\Omega_0}\delta_{ij}\delta_{k\ell}+\frac{2}{3}\frac{E_K^0}{\Omega_0}(\delta_{ik}\delta_{j\ell}+\delta_{i\ell}\delta_{jk}).
\eeq

If we consider only the kinetic part of the energy we obtain the free electron gas, and it is shown that the unperturbed system is at a pressure $\mathcal{P}$ defined as 
\beq
\sigma_{ij}^0=- \mathcal{P}\delta_{ij},
\eeq
which, according to equation \eqref{eq:sigmaP0}, is identical to that of the free electron gas derived from Fermi-Dirac statistics. However, equation \eqref{eq:Cfree} shows that the electron gas is not a fluid, but a solid with Lam\'e coefficients
\beq
\label{eq:lame}
\lambda_S=\mu_S=\mathcal{P}
\eeq
which actually gives the same bulk modulus of the free electron gas
\beq
B=\lambda_S+\frac{2}{3}\mu_S=\frac{5}{3}\mathcal{P},
\eeq
although the nature of the gas is not that of a fluid but of a solid, due to the presence of the shear modulus $\mu_S$. This result, already obtained in references \cite{tokatly1999hydrodynamic,conti1999elasticity} using kinetic arguments, suggest that the hydrodynamic model, which ignores this shear modulus, should be revisited. 

The electron gas moves actually as an isotropic solid, and its equation of motion should be that of elastodynamics. In the following two subsections the interaction terms will be added which will consist in a body force due to the single particle potential and a generalized stiffness tensor due to the two-body interaction.
\color{black}

%%%%%%%%%%%%%%%%%%%%%%%%%%%%%%%%%%%%%%%%%%%%%%%%%%%%%%%%%%%%%%%%%%%%%%%%%%%%%%%%%
\subsection{One-Particle Term}
Let us consider the contribution of the periodic potential to the elastic energy. Expanding the potential energy up to second order in the deformation we obtain
\beq
V(\fr')=V(\fr)+\fu\cdot\nabla V(\fr)+\frac{1}{2}u_iu_j\frac{\partial^2 V(\fr)}{\partial x_i\partial x_j}.
\eeq
The first term of the above expression corresponds to the unperturbed system, the second and the third ones contribute therefore to the elastic energy. If we take the expected value of the energy we obtain,
\beq
E=-\nabla\cdot\fu \braket{V}+\frac{1}{2}u_iu_j\braket{V_{ij}}
\eeq
where we have integrated by parts for the first term but not in the second one. The energy density is therefore
\beq
\mathcal{E}\approx -\frac{\braket{V}}{\Omega_0}\epsilon_{\ell \ell}+\frac{\braket{V}}{\Omega_0}\epsilon_{\ell \ell}\epsilon_{jj}+\frac{1}{2}u_iu_j \frac{\braket{V_{ij}}}{\Omega_0}
\eeq
where we have considered as well the variation of the volume $\Omega$ with the strain.  We see then that the effect of the single particle potential is to add a quantity $\braket{V}/\Omega_0$ to the equilibrium pressure $\mathcal{P}$ and to the Lam\'e parameter $\lambda_S$, while the shear modulus is not affected. Additionally, a term proportional to the square of the deformation appears, which will be the responsible of a body force density $\bm{f}=-\nabla_{\fu}\mathcal{E}$, being
\beq
f_i=-\frac{\braket{V_{ij}}}{\Omega_0}u_j.
\eeq
This term is clearly the responsible of the local dielectric function, and we would recover the classical oscillator model if we neglect the contribution of the non-local strain contribution.
%%%%%%%%%%%%%%%%%%%%%%%%%%%%%%%%%%%%%%%%%%%%%%%%%%%%%%%%%%%%%%%%%%%%%%%%%%%%%%%%%
\subsection{Two-Body Interaction}
The two-body interaction is traditionally introduced in the theory of phonons to derive the acoustic equation of motion, however in our case the two-body interaction takes place between electrons, since we are interested in the optical regime where the nucleus will remain at rest. If we assume that the deformation is small we have that 
\beq
|\fr_\alpha'-\fr_\beta'|\approx|\fr_\alpha-\fr_\beta|+\epsilon_{ij}\frac{(r_{i\alpha}-r_{i\beta})(r_{j\alpha}-r_{j\beta})}{|\fr_\alpha-\fr_\beta|}.
\eeq
Then, using $r_{\alpha\beta}=|\fr_\alpha-\fr_\beta|$, the two body potential is
\beq
U(r_{\alpha\beta}')\approx U(r_{\alpha\beta})+\epsilon_{ij}\frac{\partial U}{\partial \epsilon_{ij}}+\frac{1}{2}\epsilon_{ij}\epsilon_{k\ell}\frac{\partial^2 U}{\partial \epsilon_{ij}\partial\epsilon_{k\ell}}
\eeq
with
\beq
\frac{\partial U}{\partial \epsilon_{ij}}=\frac{\partial U}{\partial r_{\alpha\beta}}\frac{\partial r_{\alpha\beta}}{\partial\epsilon_{ij}}=\frac{\partial U}{\partial r_{\alpha\beta}}\frac{(\fr_{\alpha\beta}\otimes\fr_{\alpha\beta})_{ij}}{r_{\alpha\beta}}
\eeq
and
\beq
\frac{\partial^2 U}{\partial \epsilon_{k\ell}\partial\epsilon_{ij}}=\frac{\partial^2 U}{\partial r_{\alpha\beta}^2}\frac{\partial r_{\alpha\beta}}{\partial\epsilon_{ij}}\frac{\partial r_{\alpha\beta}}{\partial\epsilon_{k\ell}}.
\eeq
The two-body interaction contributes therefore with a fourth-rank tensor to the stiffness of the system. Since this contribution has to be averaged through the unit cell and electrons are assumed to be Bloch wave functions, we expect this tensor to have the symmetries of the lattice.

%%%%%%%%%%%%%%%%%%%%%%%%%%%%%%%%%%%%%%%%%%%%%%%%%%%%%%%%%%%%%%%%%%%%%%%%%%%%%%%%%
\section{Elastodynamic Formulation}
\label{sec:EF}
In the previous section we have shown that the elastic energy density of a deformed ensemble of electrons is given by
\beq
\mathcal{E}=\sigma_{k\ell}^0\epsilon_{k\ell}+\frac{1}{2}C_{ijk\ell}\epsilon_{ij}\epsilon_{k\ell}+\frac{1}{2}\kappa_{ij}u_iu_j,
\eeq
where the intrinsic stress of the system is
\beq
\sigma_{ij}^0=-\left(\frac{2}{3}\frac{E_K^0}{\Omega_0}+\frac{\braket{V}}{\Omega_0}\right)\delta_{ij}+\frac{1}{2}\sum_{\alpha\neq\beta}\braket{\frac{\partial U(r_{\alpha\beta})}{\partial \epsilon_{ij}}}
\eeq
and the stiffness and body force tensors are given by
\beq
\begin{split}
C_{ijk\ell}=\left(\frac{2}{3}\frac{E_K^0}{\Omega_0}+\frac{\braket{V}}{\Omega_0}\right)\delta_{ij}\delta_{k\ell}+\\
\frac{2}{3}\frac{E_K^0}{\Omega_0}(\delta_{ik}\delta_{j\ell}+\delta_{i\ell}\delta_{jk})+\sum_{\alpha\neq\beta}\braket{\frac{\partial^2 U(r_{\alpha\beta})}{\partial \epsilon_{k\ell}\partial\epsilon_{ij}}}
\end{split}
\eeq
and
\beq
\kappa_{ij}=\braket{\frac{\partial^2 V(\fr)}{\partial x_i\partial x_j}},
\eeq
respectively. We have assumed however a time-independent deformation, for which the above energy is uniquely the potential energy of the system. To derive the classical equation of motion fo the deformation we have to assume that a kinetic energy term due to the time-variation of $\fu$ will appear, so that the Lagrangian density of the ensemble can be expressed as
\beq
\mathcal{L}=\frac{1}{2}\rho_M\left(\frac{\partial \fu}{\partial t}\right)^2-\sigma_{k\ell}^0\epsilon_{k\ell}-\frac{1}{2}C_{ijk\ell}\epsilon_{ij}\epsilon_{k\ell}-\frac{1}{2}\kappa_{ij}u_iu_j
\eeq
where $\rho_M$ is the mass density of the electron solid. Assuming a constant intrinsic stress, results in the following equation of motion\cite{goldstein2002classical}
\beq
\rho_M\frac{\partial^2 u_i}{\partial t^2}=\frac{\partial}{\partial x_j}\left(C_{ijk\ell }\frac{\partial u_k}{\partial x_\ell}\right)-\kappa_{ij}u_j+F_i^e.
\eeq
Since the induced polarization is $-\rho_e\fu$ and the external force is due to the electric field, then $F_i^e=\rho_eE_i$, where $\rho_e$ is the charge density due to electrons, the above equation can be expressed as the system of equations
\subeqs{
\rho_M\frac{\partial^2 \fP}{\partial t^2}&=\nabla\cdot\sigma-\kappa\fP+\rho_M\eps_0\omega_P^2\fE \label{eq:pforce},\\
\sigma&=\bm{C}:\nabla\fP\label{eq:Hooks},
}
where we have used the definition of the plasma frequency $\eps_0\omega_P^2=\rho_e^2/\rho_M$. 

Finally, dissipation can be added phenomenologically to this model. The problem of dissipation is complex and an accurate description is beyond the objective of the present work. It will be however considered in a phenomenological way. The local mechanism for dissipation is usually introduced by means of Ohms law, in which a finite conductivity appears so that an induced current proportional to the electric field is excited.

However, dissipation in elastodynamics occurs in a different way, since for both fluids and solids it is due to forces proportional to the time derivatives of the strain. In our model, due to the presence of energy terms proportional to both the square of the strain and the deformation, we will assume that the two types of dissipation could appear, i.e., due to time derivatives of the strain and the deformation, what means that a local dissipative force $\bm{F}_\gamma=-\gamma\partial_t\fP$ will appear in equation \eqref{eq:pforce} and a viscous stress $\sigma_\eta=\bm{\eta}:\partial_t\nabla \fP$ will be added in equation \eqref{eq:Hooks}, obtaining similar terms to those derived in \cite{tokatly1999hydrodynamic,de2018viscoelastic}.

In summary, the equation of motion of the electromagnetic field inside matter when no sources are present can now be described by means of the following set of equations, 
\subeqs{
\nabla\times\fE&=-\frac{\partial\fB}{\partial t} \label{eq:rotE2} \\
\nabla\times\fB&=\mu_0\eps_0\frac{\partial\fE}{\partial t} +\mu_0\frac{\partial\fP}{\partial t}  \label{eq:rotB2}\\
\nabla\cdot\fB&=0\label{eq:divB2}\\
\nabla\cdot(\eps_0\fE+\fP)&=0 \label{eq:divE2}\\
\rho_M\frac{\partial^2 \fP}{\partial t^2}&=\nabla\cdot\sigma-\kappa\fP-\gamma \frac{\partial\fP}{\partial t} +\rho_M\eps_0\omega_P^2\fE \label{eq:P}\\
\sigma&=\bm{C}:\nabla \fP+\bm{\eta}:\partial_t\nabla \fP \label{eq:Omega} 
}
where we have used in equation \eqref{eq:divE2} the continuity equation
\beq
\nabla\cdot\fP+\rho_i=0.
\eeq

We can combine equations \eqref{eq:rotE2} and \eqref{eq:rotB2} in the usual way to have a second order partial differential equation relating $\fP$ and $\fE$. Similarly, we can combine equation \eqref{eq:Omega} and \eqref{eq:P} and we will have another second order partial differential equation relating $\fP$ and $\fE$. Combining these two we will have a fourth order partial differential equation, which in turn means that the eigenvalue problem for the fields will be a $3\times3$ matrix equation with a fourth power in the wavenumber, so that  in principle we will have three polarizations times the four solutions for the wavenumbers, i.e., a total of twelve modes, but the restrictions due to the equations \eqref{eq:divB2} and \eqref{eq:divE2} reduces by two the number of modes, then the most general solution of equations \eqref{eq:rotE2} to \eqref{eq:Omega} in a homogeneous material will consist in ten propagating modes, but with double degeneracy due to reciprocity. This means that, at an interface between two different materials, we will need to match five modes at each side, so that we will need ten boundary conditions. Equations \eqref{eq:rotE2} to \eqref{eq:Omega} will allow us to derive these ten boundary conditions in a  natural way, as will be discussed in the following section.

However, in some practical situations one of the materials can be a local material, as is the case of vacuum, local dielectrics or metals. In these situations, we will have less modes than boundary conditions, so that we will need to find the boundary conditions that are no longer satisfied. This is the opposite to the traditional approach in non-local electrodynamics, in which a great effort has been done trying to find additional boundary conditions. Clearly this approach is more efficient, since going from the most general situation to particular cases is always easier than beginning with a particular case and trying to find the most general one, as will be seen in next section.

%%%%%%%%%%%%%%%%%%%%%%%%%%%%%%%%%%%%%%%%%%%%%%%%%%%%%%%%%%%%%%%%%%%%%%%%%%%%%%%%%%%%%%%%%%%
\subsection{Boundary Conditions}
\label{sec:BC}
The set of equations \eqref{eq:rotE2} to \eqref{eq:Omega} define the evolution of the electromagnetic and polarization fields. Boundary conditions arise in a natural way in this description, since at an interface (flat or infinitesimally flat) the parallel wavenumber is a conserved quantity, and we can make in these equations the substitution $\nabla\to \fn \partial_n+i\fK_t$. Since no singularities are allowed in any of the fields (no surface currents or fields), we impose the continuity of all the fields in front of the normal derivative $\partial_n$. It is easy to deduce then that the following conditions have to be satisfied at the boundary,
\begin{subequations}
\label{eq:BC}
\subeqs{
\left[\fE_t\right]&=0\label{eq:CE}\\
\left[\fB_t\right]&=0\label{eq:CB}\\
\left[\fn\cdot\sigma\right]&=0\label{eq:CO}\\
\left[\fP\right]&=0.\label{eq:CP}
}
\end{subequations}
where $[u]\equiv u^+-u^-$. The same conclusion could be reached with the traditional pill-box and circulation arguments. The first two equations are the well-known continuity equations of electrodynamics, the only difference is that in magnetic materials the microscopic field $\fB_t$ has to be replaced by $\fH_t$. The last two equations are identical to elastodynamics, where the continuity of the normal components of the stress tensor and the displacement vector are required. 

The electrodynamic boundary conditions provide four equations (there are two components of each transverse field), while the elastodynamic ones provide us of six (the normal component of a second rank tensor is a three-component vector, and the polarization vector has three components), therefore the above equations perfectly defines the general boundary value problem at the interface between two materials, where ten equations were needed to match the five modes excited at each material, as explained before.

It is interesting to point out that within the elastodynamic description the required boundary conditions are the continuity of the transverse components of the electric and magnetic fields, however, from equation \eqref{eq:divB} the continuity of the normal component of the magnetic field also holds, as usual, but also the continuity of the normal component of the electric field, since in equation \eqref{eq:divE} the continuity of the normal component of the polarization field also implies the continuity of the normal component of the electric field. The absence of discontinuities in the fields suggests that the elastodynamic description is more suitable for the study of nanostructured materials, where the continuous nature of the wavefunctions of the different polarization carriers has also to be imposed.%%%%%%%%%%%%%%%%%%%%%%%%%%%%%%%%%%%%%%%%%%%%%%%%%%%%%%%%%%%%%%%%%%%%%%%%%%%%%%%%%%%%%%%%%%

%%%%%%%%%%%%%%%%%%%%%%%%%%%%%%%%%%%%%%%%%%%%%%%%%%%%%%%%%%%%%%%%%%%%%%%%%%%%%%%%%%%%%%%%%%
%%%%%%%%%%%%%%%%%%%%%%%%%%%%%%%%%%%%%%%%%%%%%%%%%%%%%%%%%%%%%%%%%%%%%%%%%%%%%%%%%%%%%%%%%%
%%%%%%%%%%%%%%%%%%%%%%%%%%%%%%%%%%%%%%%%%%%%%%%%%%%%%%%%%%%%%%%%%%%%%%%%%%%%%%%%%%%%%%%%%%
\subsection{Vacuum and Local Materials}
\label{sec:BCLV}
%%%%%%%%%%%%%%%%%%%%%%%%%%%%%%%%%%%%%%%%%%%%%%%%%%%%%%%%%%%%%%%%%%%%%%%%%%%%%%%%%%%%%%%%%%
The boundary conditions derived in the previous subsection emerge in a natural way at the interface between two nonlocal materials. However, in many practical situations, one of the materials might be vacuum or a local material, in the sense that the effects of the stiffness tensor might be neglected. It will be also interesting to determine the conditions for the consideration of non-local effects, that is, the conditions under which the material cannot be considered a local material and this more complex theory has to be applied. We expect this limit happens for low frequencies, but these conditions will be derived later on. In this subsection we will just consider what happens at an interface between a non-local material (solid) and a local one, which can be either vacuum or a local dielectric with finite conductivity, since, in terms of the number of solutions, all these materials are identical.

When one of the interfaces is vacuum or a local material the number of modes to match reduces from ten to seven, since now we have the five modes of the solid material plus the two polarizations allowed in local electrodynamics. Therefore, we will need as well seven boundary conditions for this interface.

Let us consider the local material first. We can assume that this situation will happen when no free or nearly free electrons are allowed (insulator), so that the stiffness component due to the kinetic term cancels, and also when the two-body interaction between electrons is negligible, due to the fact that the restoring force constant $\kappa$ is so high that electrons remain bounded around the nucleus. The material is then polarizable only locally. At the interface, the electrons traveling through the solid find a very tinny potential barrier, so that the electronic wavefunction will be continuous across the interface and, consequently, the polarization vector. However, it is clear that the free electrons from the solid will be able to apply a force at the boundary, but not the electrons at the local side, since the stress tensor is zero in the local material. 
In the case of local metals, these can be seen as those materials where the local dissipation term $\gamma$ is high, so that the mean free path of the electrons is small. The effect is that the behaviour of the material is local, since electrons are similarly ``trapped'' in a finite region, so that we will have a similar situation as described before.

Thus, the stress tensor can be discontinuous at this interface, and it is this boundary condition which is released here. We have then the following seven boundary conditions at the local-solid interface:
\begin{subequations}
\label{eq:BCL}
\subeqs{
\left[\fE_t\right]&=0\label{eq:BCL1}\\
\left[\fB_t\right]&=0\label{eq:BCL2}\\
\left[\fP\right]&=0\label{eq:BCL3}.
}
\end{subequations}
It is tempting now to define vacuum as a ``material'' in which $\kappa$ is zero and also the stiffness tensor $C$. However, this would be a wrong picture of the vacuum-solid interface. Vacuum is indeed a region non-accesible for the electrons of the solid, in the sense that these are bounded on its volume, we can consider therefore that it is the same situation as before but with the potential barrier found by the electrons in the solid being infinite, then it is the case when $\kappa\to\infty$ so that we have to impose the cancelation of the wavefunction at its boundary and, equivalently, the cancelation of the polarization vector. This is indeed the same boundary condition as \eqref{eq:BCL3}, since in vacuum the polarization vector is zero and its continuity implies the cancelation at the boundary.

This is equivalent to the mechanically rigid body, in which we impose that the displacement of the surface be zero, so that $\fP=0$, while we are able to apply a force on its surface, meaning that $\sigma$ will be different than zero there. Then,  boundary conditions will be
\begin{subequations}
\label{eq:BCV}
\subeqs{
\left[\fE_t\right]&=0\label{eq:BCV1}\\
\left[\fB_t\right]&=0\label{eq:BCV2}\\
\fP&=0\label{eq:BCV3}.
}
\end{subequations}
The above boundary conditions are the ones used by Pekar\cite{pekar1958theory}, using the argument that the excitonic wavefunctions should be zero at a vacuum interface. In section \ref{sec:planar} we will numerically show the above boundary conditions as a limiting case of $\kappa\to\infty$, and we will show numerically the transition from a non-local material to a local one.

The properties of plasmonic materials, when described by means of the hydrodynamical model\cite{melnyk1970theory}, implies only one additional mode, the longitudinal one, thus it is not possible to cancel the three polarization components, and only the normal one is imposed . However, the justification of the third ``additional boundary condition'' is typically done by suggesting that since no charge is leaving the surface we need that the normal component of the current and, therefore, of the polarization vector, has to be zero there. However this is a tricky argument, since in the local description of metals with finite conductivity the normal component of the electric current is different than zero, and it does not mean that charges are leaving the surface, it means that we have charges on the surface. This is what happens indeed during the non-stationary regime of the charge of a capacitor.

In this unified picture of spatial dispersion, these boundary conditions arise in a more natural and general way, and we can assume that the normal component of the polarization is continuous in general and it cancells at the solid-bacuum interface, because vacuum is a rigid body and then its surface cannot be displaced.

%%%%%%%%%%%%%%%%%%%%%%%%%%%%%%%%%%%%%%%%%%%%%%%%%%%%%%%%%%%%%%%%%%%%%%%%%%%%%%%%%%%%%%%%%%

\subsection{Final remarks about generalized boundary conditions}
\label{sec:discussion}
It is the current view within the community working on non-local effects that boundary conditions cannot be uniquely established, mostly because they depend on the microscopic properties of each material and  interface. Although this is certainly true for local materials, we propose to deal with the ``microscopic'' dependence of boundary conditions by including a few macroscopic quantities which help to establish the type of boundary in each interface. This approach essentially includes the strain tensor $\sigma$, and deduces a local relationship with the induced strain $\nabla\fP$. With this simple hypothesis several response models are found. For instance, the different boundary conditions discussed in \cite{halevi1984generalised} or \cite{kupresak2018comparison} can be derived by just changing the values of the stiffness tensor $\bm{C}$, local resonance $\omega_R$, plasma resonance $\omega_P$ and inertia $\rho_M$ on one side and the other of the interface, however here boundary conditions are derived within the framework of the field equations, which is a more rigorous procedure. The statement  ``boundary conditions depend on the microstructure of the material'' acquires a precise meaning in this context, since, obviously, the different values and symmetries of these parameters depend on the microstructure of the material.
Limiting situations, like perfect conductors, are perfectly defined as particular situations where the boundary conditions are different. 

The present formulation renounces to the design of the constitutive parameters, it can be left behind to other domains of physics, and within the framework of this theory we can just study the physical properties of these materials assuming at least the order of magnitude of these parameters, but also imagine new phenomena and devices and guess which values of these constants would be required. Obviously this is not the most general theory possible but, as we have found in the literature, it is a big step in the domain of nanophotonics since it unifies in just one model several materials widely studied.

The elastodynamic model is ``closed'', in the sense that we do not need additional physical considerations to properly define boundary conditions, while it is true that boundary conditions when spatial dispersion is present requires of the knowledge of the microstructure of matter, it is also true that phenomenological theories have been derived in all domains of physics, where the different limiting values of the constitutive parameters can define different materials and responses, and this description seems a good step forward towards the general understanding of spatial dispersion at the nanoscale. 

%%%%%%%%%%%%%%%%%%%%%%%%%%%%%%%%%%%%%%%%%%%%%%%%%%%%%%%%%%%%%%%%%%%%%%%%%%%%%%%%%%%%%%%%%%
\section{Solution in isotropic media}
\label{sec:ISO}
Although our approach is general and can be applied to any anisotropic material, we will illustrate the application of this method 
with the isotropic case.

Let us assume that the fields have time-harmonic dependence of the form $\exp(-i\omega t)$, in this case the constitutive equation \eqref{eq:Omega} defines a complex stiffness tensor which in isotropic media has only two independent components, then we have
\beq
C_{ij m\ell}-i\omega\eta_{ijm\ell}=\alpha\delta_{ij}\delta_{m\ell}+\beta (\delta_{im}\delta_{j\ell}+\delta_{jm}\delta_{i\ell}),
\eeq
where $\alpha$ and $\beta$ are complex parameters of the form $\alpha=\alpha_R-i\omega\alpha_I$.  The stress tensor is in this case given by
\beq
\label{eq:Oiso}
\sigma_{ij}=\alpha \nabla\cdot \fP \delta_{ij}+\beta (\partial_i P_j+\partial_j P_i),
\eeq
and the equation of motion for $\fP$ becomes
\beq
\label{eq:forcedEL}
(\alpha+\beta)\nabla \nabla\cdot\fP+\beta\nabla^2\fP-\Gamma\fP=-\rho_M\eps_0\omega_P^2\fE,
\eeq
where we have defined the frequency-dependent $\Gamma$ constant as $\Gamma=\rho_M(\omega_R^2-\omega^2)-i\gamma\omega$, with $\kappa=\rho_M\omega^2_R$. The above equation is equivalent to
\subeqs{
\left[(\alpha+2\beta)\nabla^2-\Gamma \right]\nabla\cdot\fP&=-\rho_M\eps_0\omega_P^2\nabla\cdot\fE,\\
\left[\beta\nabla^2-\Gamma \right]\nabla\times\fP&=-\rho_M\eps_0\omega_P^2\nabla\times\fE,
}
%\beq
 %\fP(\fK,\omega)=-\frac{\eps_0}{\gamma k_0^2-\beta k^2}\fE_T(\fK,\omega)-\frac{\eps_0}{\gamma k_0^2-(\alpha+2\beta)k^2}\fE_L(\fK,\omega)
 %\eeq
which in Fourier space gives
 \subeqs{
\fK\cdot\fP&=\frac{\rho_M\eps_0\omega_P^2}{\Gamma+(\alpha+2\beta)k^2}\fK\cdot\fE, \label{eq:PEk1}\\
\fK\times\fP&=\frac{\rho_M\eps_0\omega_P^2}{\Gamma+\beta k^2}\fK\times\fE \label{eq:PEk2}.
 }
 which allow us to solve for the polarization as
 \beq
 \fP=\frac{\rho_M\eps_0\omega_P^2}{\Gamma+\beta k^2}\left[\fE-\frac{\alpha+\beta}{\Gamma+(\alpha+2\beta)k^2}\fK\cdot\fE\fK\right]
 \eeq
 If we divide the electric field in a polarization parallel ($\parallel$) and perpendicular ($\perp$) to the wavevector, 
 \beq
 \fE=\fE_\parallel+\fE_\perp
 \eeq
 it is clear that we get
 \beq
 \fP=\epsilon_0\chi_\parallel\fE_\parallel+\epsilon_0\chi_\perp\fE_\perp
 \eeq
 with
 \subeqs{
\chi_\parallel(\omega,k) &=\frac{\rho_M\omega_P^2}{\Gamma+(\alpha+2\beta)k^2} \label{eq:chiPar}\\
\chi_\perp(\omega,k) &=\frac{\rho_M\omega_P^2}{\Gamma+\beta k^2} \label{eq:chiPer}.
 }

The dispersion relations and polarizations supported by the material can be found Fourier transforming equations \eqref{eq:rotB2} and \eqref{eq:rotE2}, so that we get 
\beq
-\fK\times\fK\times\fE=k_0^2\fE+k_0^2/\eps_0\fP,
\eeq 
or
 \beq
-\fK\cdot\fE \fK+k^2\fE=k_0^2\fE+k_0^2/\eps_0\fP,
\eeq 
from which we can solve for the longitudinal wavenumber
\beq
\label{eq:kL}
k_L^2=-\rho_M\frac{\omega_R^2+\omega_P^2-\omega^2-i\gamma\omega}{\alpha+2\beta},
\eeq
and the transverse one
\beq
\beta k_T^4-(\beta k_0^2-\Gamma)k_T^2- k_0^2(\Gamma+\rho_M\omega_P^2)=0,
\eeq
whose solutions are
\beq
%k_T^2=\frac{1}{2\beta}\left[\gamma (\omega)+\beta k_0^2\pm\sqrt{(\gamma (\omega)+\beta k_0^2)^2-4\beta k_0^2(\gamma(\omega)-1)}\right]
\label{eq:kT}
k_T^2=\frac{1}{2\beta}\left[\beta k_0^2-\Gamma\pm\sqrt{(\beta k_0^2+\Gamma)^2+4\rho_M\omega_P^2\beta k_0^2)}\right],
\eeq
which shows that for real $\beta$ and $\Gamma$, i.e., with no dissipation, the squared wavenumber $k_T^2$ is always real, but can be negative or positive, therefore we will have either a propagating wave or a non-dissipative evanescent one. 

In summary, we have a longitudinal wave with wave number $k_L$ and two transverse modes with wavenumbers $k_T$ given by the two solutions of (\ref{eq:kT}). Taking into account that each transverse mode is decomposed in two polarizations, we have a total of five modes propagating through the bulk material in each direction, as discussed before.

\subsection{The quasi-local limit}
It is interesting to see that, in the absence of dissipation, if $\omega_R^2+\omega_P^2-\omega^2<0$ the longitudinal mode is propagative, and also the two transverse modes $k_T$, since the product of the two solutions of equation \eqref{eq:kT} is
\beq
k_{T1}^2k_{T2}^2=-\rho_M\frac{\omega_R^2+\omega_P^2-\omega^2}{\beta}.
\eeq
However, if $\omega_R^2+\omega_P^2-\omega^2>0$, only one of the transverse modes is propagative, and the longitudinal mode is also evanescent. We have therefore only two propagative modes and three evanescent ones. This is the situation that will be analyzed in this work, since in terms of propagative modes the number of solutions is the same as in local materials, however the role of the evanescent modes is to keep the continuity of the fields at the interface.  Also, this situation will allow us to recover the local limit, as will be seen later, therefore we call this limit the ``quasi-local'' limit.

Then, in the limit $k_0^2\beta\to 0$, that is, the low frequency and low spatial dispersion limit, we can expand the solution for $k_T^2$ as
\beq
k_T^2\approx\frac{\beta k_0^2-\Gamma}{2\beta}\pm\frac{\beta k_0^2+\Gamma}{2\beta}\left(1+\frac{2\rho_M\omega_P^2\beta k_0^2}{(\Gamma+k_0^2\beta)^2}\right)
\eeq
which gives two physically different solutions, the propagating one (taking the plus sign in the $\pm$ term), and hereafter labeled $a1$,
\beq
\label{eq:ka1D}
k_{a1}^2\approx (1+\chi_E)k_0^2,
\eeq
and the evanescent one, labeled as $a2$,
\beq
\label{eq:ka2D}
k_{a2}^2\approx -\frac{\rho_M\omega_R^2}{\beta}\equiv-\frac{1}{\ell_0^2},
\eeq
where we have defined the static electric susceptibility $\chi_E$ as
\beq
\chi_E=\frac{\omega_P^2}{\omega_R^2-\omega^2-i\rho_M^{-1}\gamma \omega}.
\eeq 
and the characteristic length $\ell_0$ as
\beq
\ell_0=\frac{\beta}{\rho_M\omega_R^2}.
\eeq
Therefore, in this limit, the propagative solution is identical to that in a dielectric material with dielectric constant $1+\chi_E$, and quickly evanescent modes will be excited up to a distance $\ell_0$ from the surface. 

The case of conductors is slightly different. A Drude metal is found in the limit of $\omega_R\to 0$, with $\gamma=0$, in the quasi-local limit we would have, assuming that $\omega_P^2>\omega^2$, 
\beq
\label{eq:ka1Cond}
k_{a1}^2\approx (1-\frac{\omega^2_P}{\omega^2})k_0^2<0,
\eeq
while in this case
\beq
\label{eq:ka2Cond}
k_{a2}^2\approx \frac{\rho_M\omega^2}{\beta},
\eeq
  so that even below the plasma frequency this model predicts the existence of a propagating mode at the speed of the electronic shear wave $c_S^2=\beta/\rho_M$. This velocity can be estimated in the following way: since $\beta$ if of the order of the elastic constants of a solid, while $\rho_M$ is about two thousand times smaller than its mass density (the ratio between the proton and electron mass) we can estimate the speed of this wave of being about forty times faster than the speed of sound in solids. This additional mode will have a ver short wavelength, so that it will be also very quickly attenuated.

Indeed, if we consider the dissipation factor $\gamma$, we recover from $\chi_E$ the conductivity predicted by Drude model
\beq
\sigma=\frac{\sigma_0}{1-i\omega\tau},
\eeq
with $\sigma_0=\rho_e^2\tau/\rho_M$ and $\tau=\rho_M/\gamma$, as expected, while for the second mode we have
\beq
k_{a2}^2\approx\frac{\omega^2}{c_S^2}\left(1+i\frac{1}{\omega\tau }\right),
\eeq
which gives
\beq
\label{eq:k2metal}
k_{a2}\approx\frac{\omega}{c_S}+\frac{i}{2c_S\tau},
\eeq
so that the length $\ell_0$ is now approximately given by the distance covered by the electrons traveling at the speed $c_S$, i.e., the shear speed of the ensemble. It will be clearly a short distance, and the local limit will be recovered in this case taking the limit of this distance to zero, with the boundary conditions derived before, as will be demonstrated later on.
\color{black}
\subsection{The characteristic length $\ell_0$}
The term ``non-local'' has been used through this work paying attention specially to the consequences in terms of wave propagation and the number of modes to match. Although it is this analysis the relevant one in terms of the solution of boundary value problems, it is also interesting the analysis of the constitutive equations in real space, since it will help us to better understand the physical interpretation of the stiffness tensor $\bm{C}$. We will limit our discussion to the isotropic case and neglecting the longitudinal response, but a deeper analysis can be found in \cite{wubs2015classification}.

For simplicity we will assume that $\alpha=0$, so that equations \eqref{eq:PEk1} and \eqref{eq:PEk2}  defines a frequency and a wavenumber dependent electrical susceptibility
\beq
\fP=\frac{\rho_M\omega_P^2\eps_0}{\Gamma+\beta k^2}\fE \equiv \eps_0\chi_{NL}(k,\omega) \fE,
\eeq
which in real space relates the polarization and the electric field as
 \beq
 \fP(\fr,\omega)=\eps_0\iiint \chi_{NL}(\fr-\fr',\omega)\fE(\fr',\omega)dV,
 \eeq
 where the non-local susceptibility is
\beq
\chi_{NL}(\fr,\omega)=\rho_M\omega_P^2\iiint \frac{e^{i\fK\cdot\fr}}{\Gamma+\beta k^2}dV_k,
\eeq 
or, using the radial symmetry of the Fourier transform,
 \beq
 \label{eq:intchi}
 \chi_{NL}(r,\omega)=\sqrt{\frac{2}{\pi}}\frac{\rho_M\omega_P^2}{r}\int_{0}^\infty \frac{k\sin k r}{\Gamma+\beta k^2}dk.
 \eeq
 Therefore, the polarization at a given point $\fr$ is a linear combination of the electric field through all the space weighted by the non-local susceptibility $\chi_E(\fr,\omega)$.  For both $\Gamma$ and $\beta$ reals and positive (the quasi-dielectric limit), we have (see \cite{gradshteyn2014table}, section 273, equation 3)
 \beq
 \chi_{NL}(r,\omega)=\sqrt{\frac{\pi}{2}}\frac{\rho_M\omega_P^2}{\beta}\frac{e^{-r/\ell_0}}{ r}=\chi_E\sqrt{\frac{\pi}{2}}\frac{e^{-r/\ell_0}}{ \ell_0^2r},
 \eeq
 where we have defined the characteristic length $\ell_0$ as
 \beq
 \ell_0^2=\beta/\Gamma.
 \eeq
 
 The meaning of the parameter $\ell_0$ is therefore clear in this limit: a region of radius $\ell_0$ has to be taken into account to compute the polarization $\fP$, since outside this region the susceptibility $\chi_{NL}$ is negligible.
 
 Once defined the characteristic length $\ell_0$, the effects of spatial dispersion will be more or less important depending on the wavelength of the field. As will be seen later, when $\lambda>>\ell_0$ the influence region is averaged and the behaviour of the fields is similar to that of a local material, while when $\lambda\approx\ell_0$ the effects of spatial dispersion are more relevant and cannot be ignored.
 
The above discussion is valid only in the quasi-local limit, that is, when all the additional modes are evanescent and we still have only one propagative solution. 
%However, when $\Gamma$ is negative, that is, when the factor $\gamma k^2_0$ dominates, we have (see \ref{}, section 273, equation 10)
% \beq
% \chi_E(r,\omega)=\sqrt{\frac{\pi}{2}}\frac{\cos{\sqrt{\Gamma}r/\ell}}{\ell^2 r}
% \eeq
%which clearly shows that the influence length is larger than $\ell$, since now the non-local susceptibility decays slower, as $1/(\ell^2 r)$, but the exponential term became an oscillatory term, and the material is even more non-local. In this regime all the five modes are propagative modes, although it does not change anything in terms of boundary conditions, clearly the physics of the problem is much different. Since in this work we are interested in the study of these boundary conditions and their meaning as we reach the local limit, we will consider that the influence length $\ell$ has indeed the meaning explained in the quasi-dielectric limit. 
When either $\Gamma$ and $\beta$ are complex, the residue theorem has to be applied to evaluate equation \eqref{eq:intchi}, and we find that the exponential decay rate depends on the absolute value of the real part of  $\sqrt{\Gamma/\beta}$. Then, let us assume for instance that $\beta$ is real,  and $\Gamma=\rho_M(\omega_R^2-\omega^2)-i\omega \gamma$, we will have
\beq
\sqrt{\Gamma/\beta}\approx\sqrt{\frac{\rho_M(\omega_R^2-\omega^2)}{\beta}}\left(1-\frac{i\omega\delta}{2\rho_M(\omega_R^2-\omega^2)}\right).
\eeq

The above result shows that, as long as $\omega_R^2-\omega^2>0$ the characteristic length is of the order of $\ell_0$, as before, however when $\omega_R^2-\omega^2<0$ the imaginary and real parts are exchanged, and the characteristic length becomes proportional to $1/\delta$, which can be very high, depending on the value of this parameter. The same effect applies as well when $\beta$ has both real and imaginary parts, but not when it is purely imaginary. A deeper analysis of this complex situation is beyond the objective of the present work, which is only to understand how the effects of spatial dispersion affects boundary conditions, therefore in the numerical calculations we will consider only the quasi-dielectric limit, in which the characteristic length is small and we can compare results with the local limit.

\section{Spill-out of electrons}
\label{sec:SO}
In section \ref{sec:BCLV} the boundary conditions for a vacuum-solid interface where discussed and it vacuum was defined as a region non-accesible for electrons, since the potential barrier between the solid and vacuum is assumed to be infinite. It can happen however that for some materials (insulators or conductors) this potential barrier be finite, which allow the spill-out of electrons from the solid to vacuum, as has been recently observed experimentally and discussed theoretically \cite{baumberg2019extreme}. The spill-out of electrons to vacuum can also be considered within the present model by means of two approaches, which will be discussed in the following two subsections. First, we will consider the situation where a small number of electrons leave the solid and then populate vacuum with a not necessarily uniform electron density $n_0$. Later we will discuss the solution by means of the density functional theory (DFT), showing however that a more advanced approach is required in this case whose detailed solution is beyond the objective  of the present work.

\subsection{Homogeneous electron density}
Let us assume that some electrons leave the solid and they jump into vacuum, so that now vacuum is a medium with some electron density $n_0$, which can be modelled as a perfect conductor, so that in equations \eqref{eq:kL} and \eqref{eq:kT} we set $\gamma=0$ and $\omega_R=0$. Additionally, if we assume that the number of electrons populating vacuum is small, we can also assume that $\omega_P=\rho_e^2/\eps_0\rho_M \approx 0$, therefore the longitudinal wavenumber will be
\beq
k_L^2=\frac{\rho_M}{\alpha+2\beta}\omega^2
\eeq
while the two transverse modes will be, from equation \eqref{eq:ka1Cond}
\beq
k_{a1}=k_0
\eeq
and from equation \eqref{eq:ka2Cond}
\beq
k_{a2}^2=\frac{\rho_M}{\beta}\omega^2
\eeq
Therefore, for this low-populated vacuum we see that the electromagnetic field has likewise five modes, which correspond to the two transverse modes with wavenumber $k_0$ and, consequently, zero polarization, plus the three polarizations resulting from the solution of the elastodynamic equation \eqref{eq:forcedEL}, which are the two transverse modes plus the longitudinal one. We can interpret this low-populated medium as one in which the electromagnetic field is not really disturbed, so that it propagates as in vacuum, however the electrons are disturbed and they oscillate as a solid. There is obviously an electromagnetic field created by this polarization, but it travels at a very small velocity.

\subsection{Inhomogeneous electron density}
The approach proposed in the previous section can be useful and accurate in the case of small gaps between solid layers, and it requires obviously an estimation of the density of electrons $n_0$ who has left the solids and are now at equilibrium in the gap. However, the electron density for a finite potential barrier will not be in general a continuous function, since the electron could can have some complex spatial dependence $n_0=n_0(\fr)$. If the equilibrium electron density $n_0(\fr)$ is previously known, we can model the interaction by solving the field equations with the corresponding position-dependent parameters. Then, we know that for the free electron gas the $\alpha$ and $\beta$ parameters correspond to the Lam\'e constants and are given by (see equation \eqref{eq:Cfree})

\beq
\alpha(\fr)=\beta(\fr)=\mathcal{P}(\fr)
\eeq
where the quantum pressure $\mathcal{P}(\fr)$  can be computed using equation \eqref{eq:PThermo}, once the inhomogeneous equilibrium energy $E(n_0(\fr))$ is known. 

Boundary conditions remains the same but since the density of electrons $n_0(\fr)$ is a continuous function, we do not need to solve a boundary value problem, which somehow solves the problem of additional boundary conditions although closed forms expressions for the scattering or transmission of waves are unlikely. 
%%%%%%%%%%%%%%%%%%%%%%%%%%%%%%%%%%%%%%%%%%%%%%%%%%%%%%%%%%%%%%%
\subsection{Density and Current Functional Theory }
In recent works the above approach has been refined by means of the density functional theory (DFT), in which the density of electrons $n_0(\fr)$ is set as a new field variable and solved simultaneously with the field equations\cite{toscano2015resonance,ciraci2016quantum,ding2018eigenvalue}. However, the DFT has been mainly developed for a scalar external potential, and the inclusion of the vector potential $\bm{A}$ in the interaction, which implies the definition of a ``current density functional theory'' (CDFT) is not as well developed as the DFT. It is easy to see then that, considering only the density functional theory the movement of the electrons are assumed to be as a gas or a fluid, so that only they hydrodynamical model is suitable for this description, which we have discussed previously to be incomplete. We see as well how within the framework of the DFT the hydrodynamical model is as well incomplete: we need the current as a functional as well. In reference \cite{ciraci2017current} Cirac\`i applies CDFT to complement the hydrodynamical model, and he uses a viscoelastic tensor to complement the equation of motion, using the expression derived in \cite{tokatly1999hydrodynamic} but without adding the non-viscous term derived in this work. We believe that the approach by Cirac\`i plus the elasticity tensor derived in this work would represent a more accurate description than that derived within the frame work of the hydrodynamic model.

Therefore, although the spill-out of electrons is a sensitive issue within the domain of non-local plasmonics, the solutions reported so far within the framework of the DFT are only approximately correct, since the hydrodynamic model is clearly incomplete. 

We can understand the limitations of the DFT plus the hydrodynamical description in the following way. Let us assume that the ensemble of electrons moves as a fluid material. Let us assume that, in equilibrium, some of the electrons can spill-out from the solids to vacuum, and finally we have an equilibrium energy density $E(n_0(\fr))$, as well as an equilibrium electron density $n_0(\fr)$, both inhomogeneous. If we assume that the energy $E$ is a functional solely of the density $n_0(\fr)$, any deviation from the equilibrium of the system will result in a restoring force due to the gradient of a pressure field $\mathcal{P}(\fr)$, derived from equation \eqref{eq:PThermo}, thus the equation of motion of the ensemble will be
\beq
\rho_M\frac{\partial \fJ}{\partial t}=\rho_e^2\fE-\rho_e\nabla\mathcal{P}(\fr)
\eeq

However, a more general functional of the energy might imply that the energy be a function not only of the equilibrium density $n_0(\fr)$, but also of the equilibrium current $\fJ_0(\fr)$ and, consequently, the response of the ensemble to any deformation will be a restoring force due to the divergence of a stress field $\sigma_{ij}$, obtained from equation \eqref{eq:sigmaE}, thus the equation of motion will be now
\beq
\rho_M\frac{\partial \fJ}{\partial t}=\rho_e^2\fE+\rho_e\nabla\cdot\sigma(\fr)
\eeq

Therefore, the only difference between the ``classical'' hydrodynamic formulation and the use of the DFT in the calculations is the equation of state relating the pressure field and the density $n_0(\fr)$, similarly, if we employ a Hooks-like law for the stress $\sigma_{ij}$ above, we recover the elastodynamic formulation, while if we use the CDFT to obtain the stress from \eqref{eq:sigmaE}, we solve the full quantum-mechanical problem without the need of additional boundary conditions, since the only part missing is Maxwell equations and then the problem is self-contained.

If we exclude those effects related to microscopic details of the  surface, we consider that the assumption of bounded domains with  well-defined homogeneous electron densities describes the underlying  physics of non-local materials. More microscopical techniques are obviously required in some extreme  situations. However, the above discussion, show that the  hydrodynamic description, i.e., the assumption that electrons behave  like a gas in- stead of like a solid, is rather incomplete and should  be revisited
%%%%%%%%%%%%%%%%%%%%%%%%%%%%%%%%%%%%%%%%%%%%%%%%%%%%%%%%%%%%%%%%%%%%%%%%%%%%%%%%%%%%%%%%%%%
\section{Further comparison with the hydrodynamic model}
\label{sec:HE}
The theory we have presented in this work can be applied to any solid (or fluid) material, since the interaction between electrons has been considered as well and it is a matter of obtaining the right constitutive parameters (with their corresponding symmetries) to properly model the optical response at the nanoscale. However, non-local effects have been specially considered for plasmonic materials, where electrons are assumed to move freely through the conduction bands for which the hydrodynamic model has been the most widely used in the study of these materials. We consider that the present model is more accurate than the hydrodynamic model for the study of free or nearly free electrons. Although through the text these differences can be found, due to its relevance we would like to summarize them in this section.
\subsection{Shear modulus}
As stated in section \ref{sec:QS}, the hydrodynamic model assumes that only longitudinal mode exist and the equation of motion for the induced current is that of acoustics, with a bulk modulus $B$ defined by means of equation \eqref{eq:Bulk}, while the present approach shows that for the free electron gas we have both a bulk modulus $B$ and a shear modulus $\mu_S$, i.e., the equation of motion for the induced current in the free electron gas is that of elastodynamics, where the shear modulus is given by equation \eqref{eq:lame}, which shows that the relationship between the shear and bulk modulus is
\beq
\mu_S=\frac{3}{5}B
\eeq
and, consequently, they are of the same order of magnitude and we should not neglect shear waves in this description. It is remarkable that we have derived equation \eqref{eq:lame} from a perturbation of the electronic wave function while in reference \cite{tokatly1999hydrodynamic} the authors derived the same expressions using Boltzman's transport equation, what reinforces the accuracy of this result.

\subsection{Longitudinal and transverse dielectric functions}
Another remarkable argument in favour of the existence of this shear modulus is that, if taken into account, we obtain both a non-local longitudinal and transverse susceptibilities, as defined by equations \eqref{eq:chiPar} and \eqref{eq:chiPer}, while if we ignore the shear modulus, the perpendicular susceptibility becomes local and we have a non-local response only in the longitudinal component. However, this contradicts the result obtained with quantum mechanics using perturbation theory for the free electron gas, since it is shown there that the response function contains both a longitudinal and transverse components\cite{grosso2003giuseppe}. There is no reason therefore to ignore one of these components and consider only one, as the hydrodynamical model assumes, so that both terms should be present in any description. 
\subsection{Nonlocal response of S-Polarized waves}
Finally, the most interesting difference arising after including or not the shear modulus in the free electron gas is the fact that, according to the hydrodynamic description, the interaction of electromagnetic waves at any interface where the electric field be parallel to the surface (i.e., normal incidence in a plane or S polarization in a plane or cylinder) will be indistinguishable from a local response. A non-local response in these situations can only be modelled (as will be shown in next section) if the shear modulus is included in the equation of motion. The non-local response of a material is due to the movement of electrons after the electromagnetic field has excited them, there is no obvious physical reason for which this response had to be local or non-local depending on the polarization of the field if the material is isotropic, and non-locality is not only a surface effect, it is a bulk property that should not depend on the orientation of the field at the interface under consideration.

Therefore, we believe that the elastodynamic description is much more accurate than the hydrodynamic one, since some of the results predicted by the hydrodynamic description are clearly non-physical.

%%%%%%%%%%%%%%%%%%%%%%%%%%%%%%%%%%%%%%%%%%%%%%%%%%%%%%%%%%%%%%%%%%%%%%%%%%%%%%%%%%%%%%%%%%%
\color{black}
\section{Scattering by planes, cylinders and spheres}
\label{sec:SC}
In the previous sections we have seen that the hydrodynamic model is not very accurate for the description of the interatction of light and electrons. We have seen that the presence of shear forces are found when considering both full quantum mechanical equations and the kinetic theory, which additionally are consistent with our ``elastodynamic approach''. In this section we will focus then on the non-local effects that the hydrodynamical cannot predict at all, i.e., those non-local effects related with the shear forces excited in the movement of electrons.

%%%%%%%%%%%%%%%%%%%%%%%%%%%%%%%%%%%%%%%%%%%%%%%%%%%%%%%%%%%%%%%%%%%%%%%%%%%%%%%%%%%%%%%%%%%

\begin{figure}[h!]
	\centering
	\includegraphics[width=\linewidth]{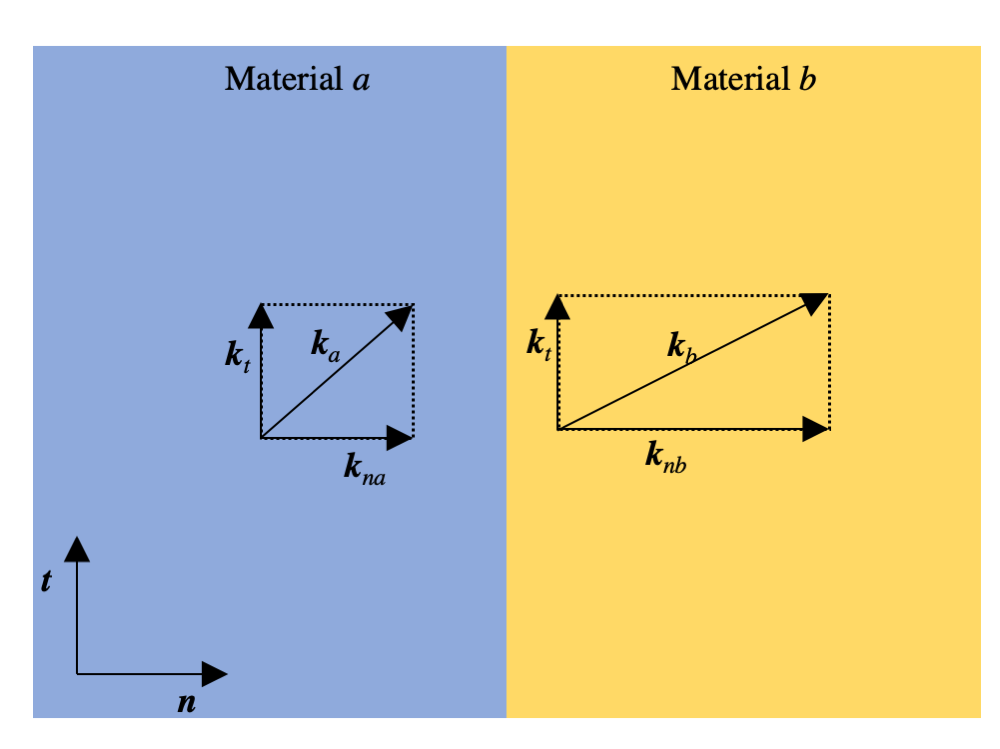}
	\caption{Schematic diagram of the interface problem discussed in the text.}
	\label{fig:schematics}
\end{figure}
%%%%%%%%%%%%%%%%%%%%%%%%%%%%%%%%%%%%%%%%%%%%%%%%%%%%%%%%%%%%%%%%%%%%%%%%%%%%%%%%%%%%%%%%%%%
\subsection{Planar interface}
\label{sec:planar}
In this section we will consider the planar interface between two solids, labeled $a$ and $b$, in order to analyze the consistency of the boundary conditions derived previously, as well as the conditions in which we recover the local limit. We will assume that material $a$ is at the left hand side of the interface and material b is at the right hand side, as shown in figure \ref{fig:schematics}.

Let us assume that we are in the quasi-dielectric limit, in which one transverse wavenumber is propagative and the other transverse wavenumber is evanescent, as well as the longitudinal one. It is convenient now to derive the relationship between the different fields involved in boundary conditions and the electric field. Thus, after performing the substitution $\nabla\to i\fK_t+\partial_n$, from equation \eqref{eq:rotE2} we have that the transverse magnetic field is given by
\beq
i\omega\fB_t=\fn\times\partial_n\fE_t+i\fK_t\times\fn E_n
\eeq
while the polarization vector is given by
\beq
\frac{k_0^2}{\eps_0} \fP=\left(k^2-k_0^2\right)\fE+(\fn\partial_n+i\fK_t)(i\fK_t\cdot \fE_t+\partial_nE_n).
 \eeq

The $ij$ element of the stress tensor $\sigma$ in an isotropic material is given in by equation \eqref{eq:Oiso}, so that the normal and transverse components, which are to be continuous, are
\subeqs{
\sigma_{nn}&=(\alpha+2\beta)\partial_nP_n+i\alpha\fK_t\cdot\fP\\
\sigma_{nt}&=\beta\partial_n \fP_t+i\beta \fK_tP_n.
}
In the next subsection we will analyze the behaviour of these fields at normal incidence, and later on we will derive the general reflection coefficients at the interface between vacuum and a solid material considering oblique incidence.

%%%%%%%%%%%%%%%%%%%%%%%%%%%%%%%%%%%%%%%%%%%%%%%%%%%%%%%%%%%%%%%%%%%%%%%%%%%%%%%%%%%%%%%%%%%
%%%%%%%%%%%%%%%%%%%%%%%%%%%%%%%%%%%%%%%%%%%%%%%%%%%%%%%%%%%%%%%%%%%%%%%%%%%%%%%%%%%%%%%%%%%
%%%%%%%%%%%%%%%%%%%%%%%%%%%%%%%%%%%%%%%%%%%%%%%%%%%%%%%%%%%%%%%%%%%%%%%%%%
%%%%%%%%%%%%%%%%%%%%%%%%%%%%%%%%%%%%%%%%%%%%%%%%%%%%%%%%%%%%%%%%%%%%%%%%%%%%%%%%%%%%%%%%%%%
\subsubsection{Normal incidence}
Let us asume that a propagating transverse wave is excited in material $a$ and propagates along the $x$ axis, so that $\fn=\fx$. Without loss of generality, we will assume that the interface is placed at $x=0$ and that the field is polarized along the $z$ direction. If the excited mode has wavevector $\fK=k_{a1}\fx$, after reflection two additional modes are excited in material $a$, corresponding to the two solutions of the dispersion equation \eqref{eq:kT}. Since there are no normal components of the fields there is no excitation of the longitudinal mode, whose discussion is left for next subsection.  Then we have, for $x<0$,
\beq
E_z=e^{ik_{a1}x}+R_1e^{-ik_{a1}x}+R_2e^{-ik_{a2}x}.
\eeq

The transverse component of the magnetic field is
\beq
\omega B_y=k_{a1}(e^{ik_{a1}x}-R_1e^{-ik_{a1}x})-k_{a2}R_2e^{-ik_{a2}x})
\eeq
and the polarization vector is parallel to the electric field, therefore
\beq
\begin{split}
\frac{k_0^2}{\eps_0}P_z=(k_{a1}^2-k_0^{2})(e^{ik_{a1}x}+R_1e^{-ik_{a1}x})+\\(k_{a2}^2-k_0^{2})R_2e^{-ik_{a2}x}.
\end{split}
\eeq

The normal component of the stress tensor $\sigma$ is zero, and the transverse one is
\beq
\begin{split}
\frac{\eps_0}{ik_0^2}\sigma_{zx}=\beta_ak_{a1}(k_{a1}^2-k_0^{2})(e^{ik_{a1}x}-R_1e^{-ik_{a1}x})-\\\beta_a k_{a2}(k_{a2}^2-k_0^{2})R_2e^{-ik_{a2}x}.
\end{split}
\eeq

For $x>0$ we have two modes excited as well, thus
\subeqs{
E_z&=T_1e^{ik_{b1}x}+T_2e^{ik_{b2}x}\\
\omega B_y&=k_{b1}T_1e^{ik_{b1}x}+k_{b2}T_2e^{ik_{b2}x}\\
\frac{k_0^2}{\eps_0}P_z&=(k_{b1}^2-k_0^{2})T_1e^{ik_{b1}x}+(k_{b2}^2-k_0^{2})T_2e^{ik_{b2}x}\\
\frac{\eps_0}{ik_0^2}\sigma_{zx}&=\beta_bk_{b1}(k_{b1}^2-k_0^{2})T_1e^{ik_{b1}x}-\beta_b k_{b2}(k_{b2}^2-k_0^{2})T_2e^{ik_{b2}x}
}

In summary, we will need to determine the value of four coefficients, and we have indeed four equations for this geometry, the continuity of $E_z, B_y,P_z$ and $\sigma_{zx}$, then the solution for the coefficients can be found after inversion of the system
\begin{widetext}
\beq
\left(\begin{matrix}
-1 & -1 & 1 & 1 \\
k_{a1} & k_{a2} & k_{b1} & k_{b2}\\
-(k_{a1}^2-k_0^{2} )& -(k_{21}^2-k_0^{2}) & k_{b1}^2-k_0^{2} & k_{b2}^2-k_0^{2} \\
\beta_a k_{a1}(k_{a1}^2-k_0^{2} )& \beta_a k_{a2}(k_{21}^2-k_0^{2}) & \beta_b k_{b1}(k_{b1}^2-k_0^{2}) & \beta_b k_{b2}(k_{b2}^2-k_0^{2} )
\end{matrix}\right)
\left(\begin{matrix}
R_1\\
R_2\\
T_1\\
T_2
\end{matrix}\right)=
\left(\begin{matrix}
1 \\
k_{a1}\\
(k_{a1}^2-k_0^{2}) \\
\beta_a k_{a1}(k_{a1}^2-k_0^{2})
\end{matrix}\right).
\eeq
\end{widetext}

The above equations show that the interface problem is well defined once we have defined the two transverse wavenumbers (at normal incidence there is no excitation of the longitudinal mode) and the $\beta$  coefficient. We will see now how the above equations, derived using equations \eqref{eq:BC}, can degenerate in equations \eqref{eq:BCL} as the limit of low characteristic length $\ell_0$. 

\begin{figure}[h!]
	\centering
	\includegraphics[width=\linewidth]{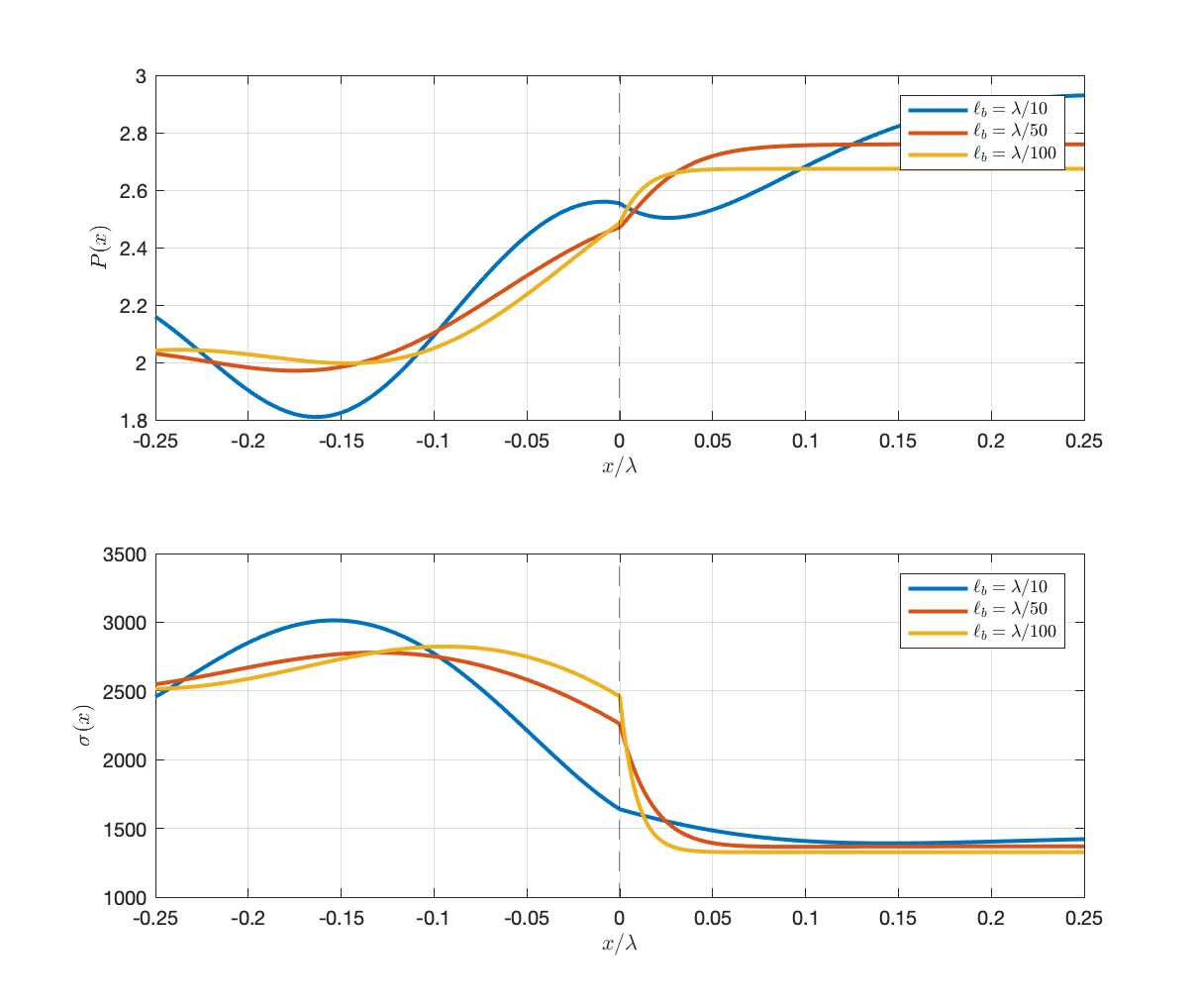}
	\caption{Polarization vector (upper panel) and stress field (lower panel) as a function of $x$ near an interface solid-solid for several values of the characteristic length $\ell_b$ of material $b$. We see how in the transition $\ell_b\to0$ material $b$ becomes a local material and the stress tensor presents a discontinuity, as predicted by the theory (see text for further details).}
	\label{fig:StoD}
\end{figure}

Figure \ref{fig:StoD} shows the polarization $P(x)$ (upper panel) and the stress tensor $\sigma_{xz}$ as a function of $x$ along an interface between two solid materials. Material $a$ is chosen so that $\beta_a=3$, $k_{a1}^2=(1+\chi_{Ea})k_0^2$ and $k_{a2}^2=-1/\ell_{a}^2$, with $\chi_{Ea}=2$ and $\ell_a=\lambda/10$. Similarly, material $b$ is selected as $\beta_b=1$, $k_{b1}^2=(1+\chi_{Eb})k_0^2$ and $k_{b2}^2=-1/\ell_{b}^2$, with $\chi_{Eb}=3$ and $\ell_b=\lambda/10,\lambda/50$ and $\lambda/100$. We see that both solids are non-local but we have analyzed the evolution of material $b$ towards a local material, showing therefore that in this transition the polarization vector remains continuous while the stress tensor presents a discontinuity, as required by boundary conditions \eqref{eq:BCL}.

%%%%%%%%%%%%%%%%%%%%%%%%%%%%%%%%%%%%%%%%%%%%%%%%%%%%%%%%%%%%%%%%%%%%%%%%%%%%%%%
%%%%%%%%%%%%%%%%%%%%%%%%%%%%%%%%%%%%%%%%%%%%%%%%%%%%%%%%%%%%%%%%%%%%%%%%%%%%%%%
%%%%%%%%%%%%%%%%%%%%%%%%%%%%%%%%%%%%%%%%%%%%%%%%%%%%%%%%%%%%%%%%%%%%%%%%%%%%%%%
%%%%%%%%%%%%%%%%%%%%%%%%%%%%%%%%%%%%%%%%%%%%%%%%%%%%%%%%%%%%%%%%%%%%%%%%%%%%%%%

\begin{figure}[h!]
	\centering
	\includegraphics[width=\linewidth]{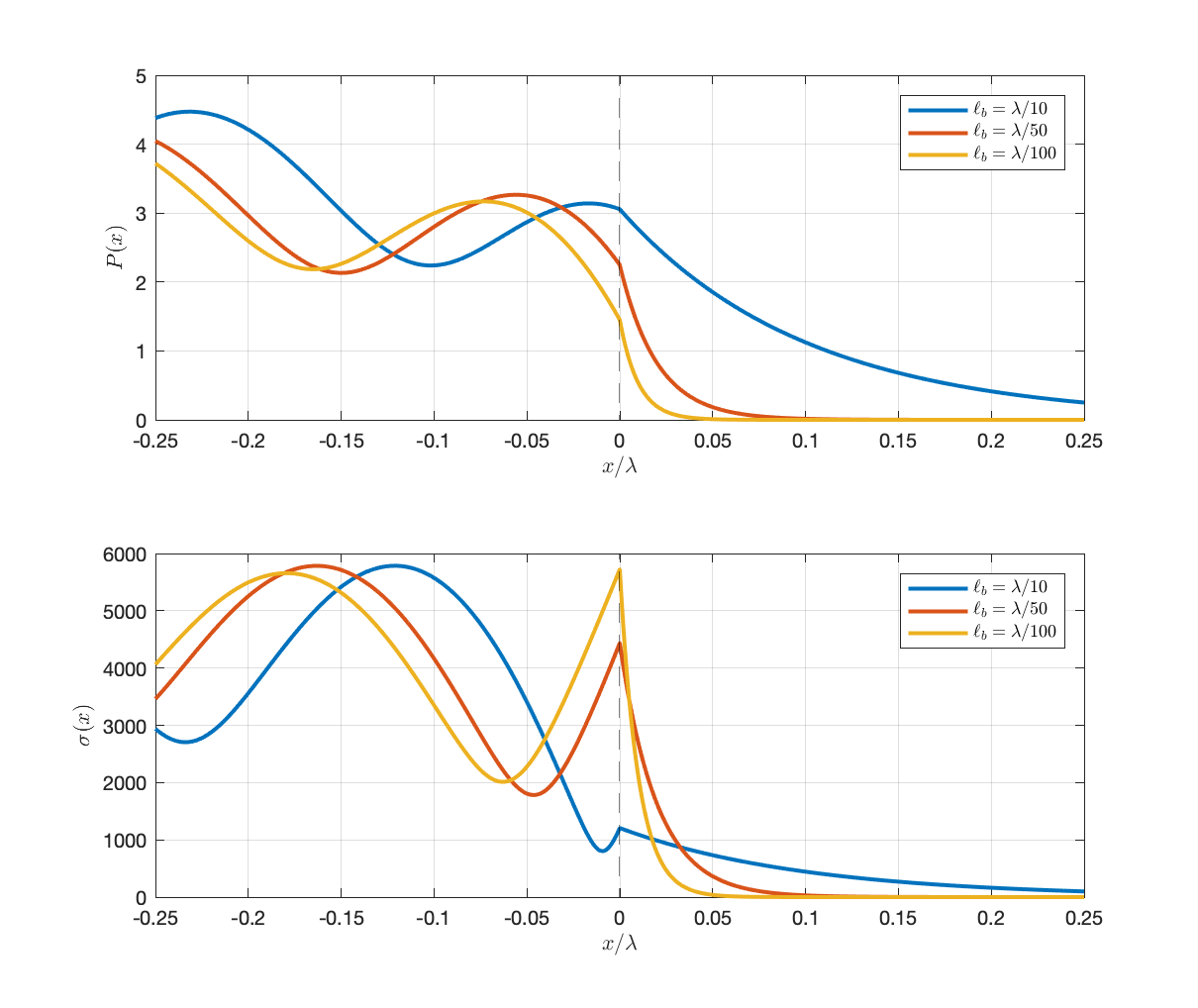}
	\caption{Same system as described in figure \ref{fig:StoD} but now material $b$ converges towards vacuum.}
	\label{fig:StoV}
\end{figure}

Figure \ref{fig:StoV} shows the same situation as before but now we set $\chi_{Eb}$ equal to 0, so that in the limit of $\ell_b\to0$ material $a$ converges towards vacuum. We see that, as expected for vacuum, both the polarization and the stress fields cancels, however the polarization remains continuous (zero) at the interface while the stress presents a discontinuity. Once more this agrees with boundary conditions \eqref{eq:BCV}.

%%%%%%%%%%%%%%%%%%%%%%%%%%%%%%%%%%%%%%%%%%%%%%%%%%%%%%%%%%%%%%%%%%%%%%%%%%%%%%%
%%%%%%%%%%%%%%%%%%%%%%%%%%%%%%%%%%%%%%%%%%%%%%%%%%%%%%%%%%%%%%%%%%%%%%%%%%%%%%%
%%%%%%%%%%%%%%%%%%%%%%%%%%%%%%%%%%%%%%%%%%%%%%%%%%%%%%%%%%%%%%%%%%%%%%%%%%%%%%%
%%%%%%%%%%%%%%%%%%%%%%%%%%%%%%%%%%%%%%%%%%%%%%%%%%%%%%%%%%%%%%%%%%%%%%%%%%%%%%%

\begin{figure}[h!]
	\centering
	\includegraphics[width=\linewidth]{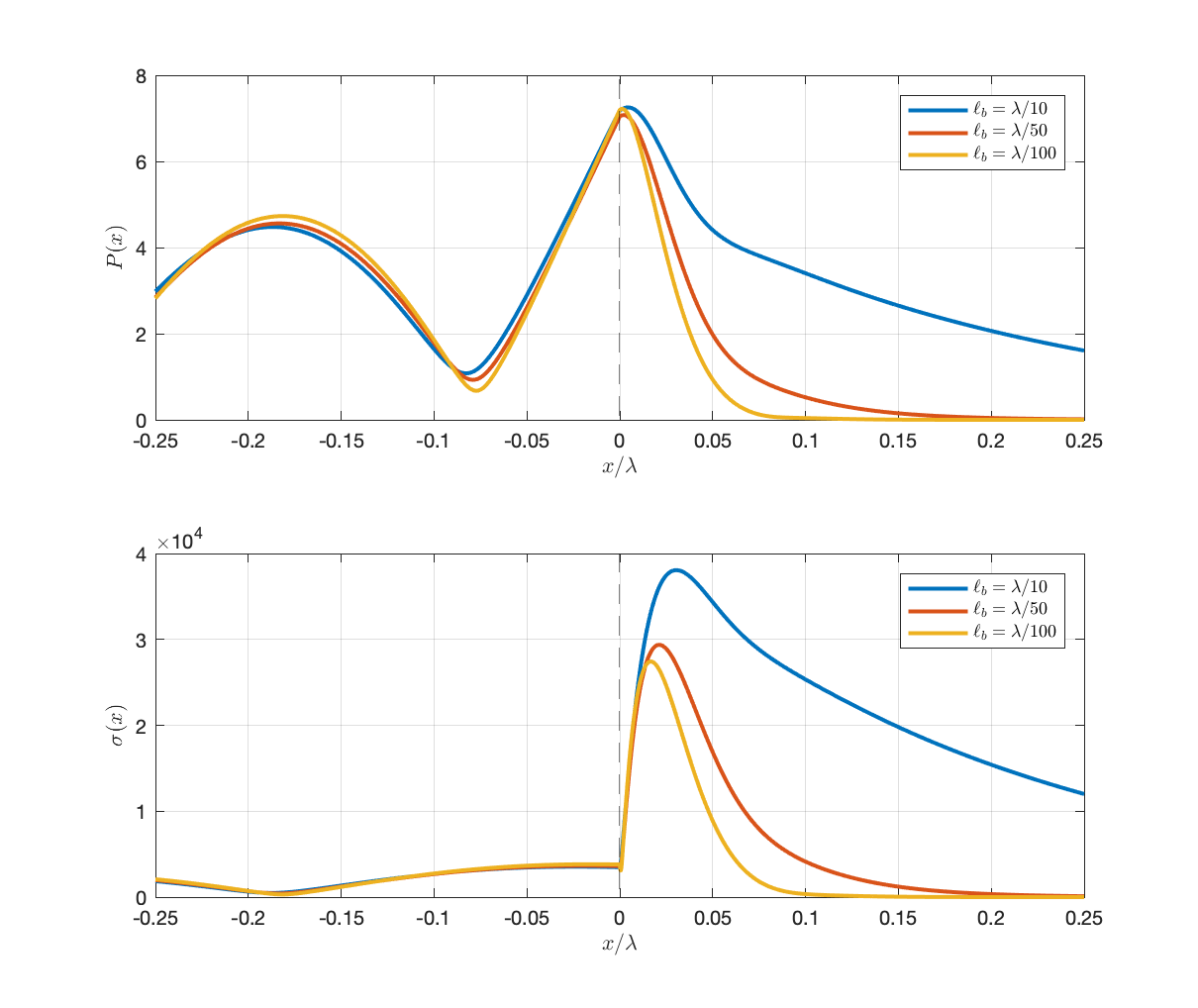}
	\caption{Same situation as in figure \ref{fig:StoD} but now material $b$ is a metal with $\chi_{Eb}=-100$ and  $k_{b2}=k_S+i/\ell_{b}$.}
	\label{fig:StoM}
\end{figure}

Figure \ref{fig:StoM} shows the same situation but now we set $\chi_{Eb}=-100$, which corresponds to a metalic material. The meaning of the characteristic length $\ell_b$ is similar, as shown in equation \eqref{eq:k2metal}, but now we add the contribution of the short wave to $k_{b2}$, so that we have $k_{b2}=k_S+i/\ell_{b}$, with $k_S=10k_0$ and $\ell_b$ having the same range as before. 
%%%%%%%%%%%%%%%%%%%%%%%%%%%%%%%%%%%%%%%%%%%%%%%%%%%%%%%%%%%%%%%%%%%%%%%%%%%%%%%
%%%%%%%%%%%%%%%%%%%%%%%%%%%%%%%%%%%%%%%%%%%%%%%%%%%%%%%%%%%%%%%%%%%%%%%%%%%%%%%
%%%%%%%%%%%%%%%%%%%%%%%%%%%%%%%%%%%%%%%%%%%%%%%%%%%%%%%%%%%%%%%%%%%%%%%%%%%%%%%
%%%%%%%%%%%%%%%%%%%%%%%%%%%%%%%%%%%%%%%%%%%%%%%%%%%%%%%%%%%%%%%%%%%%%%%%%%%%%%%

\begin{figure}[h!]
	\centering
	\includegraphics[width=\linewidth]{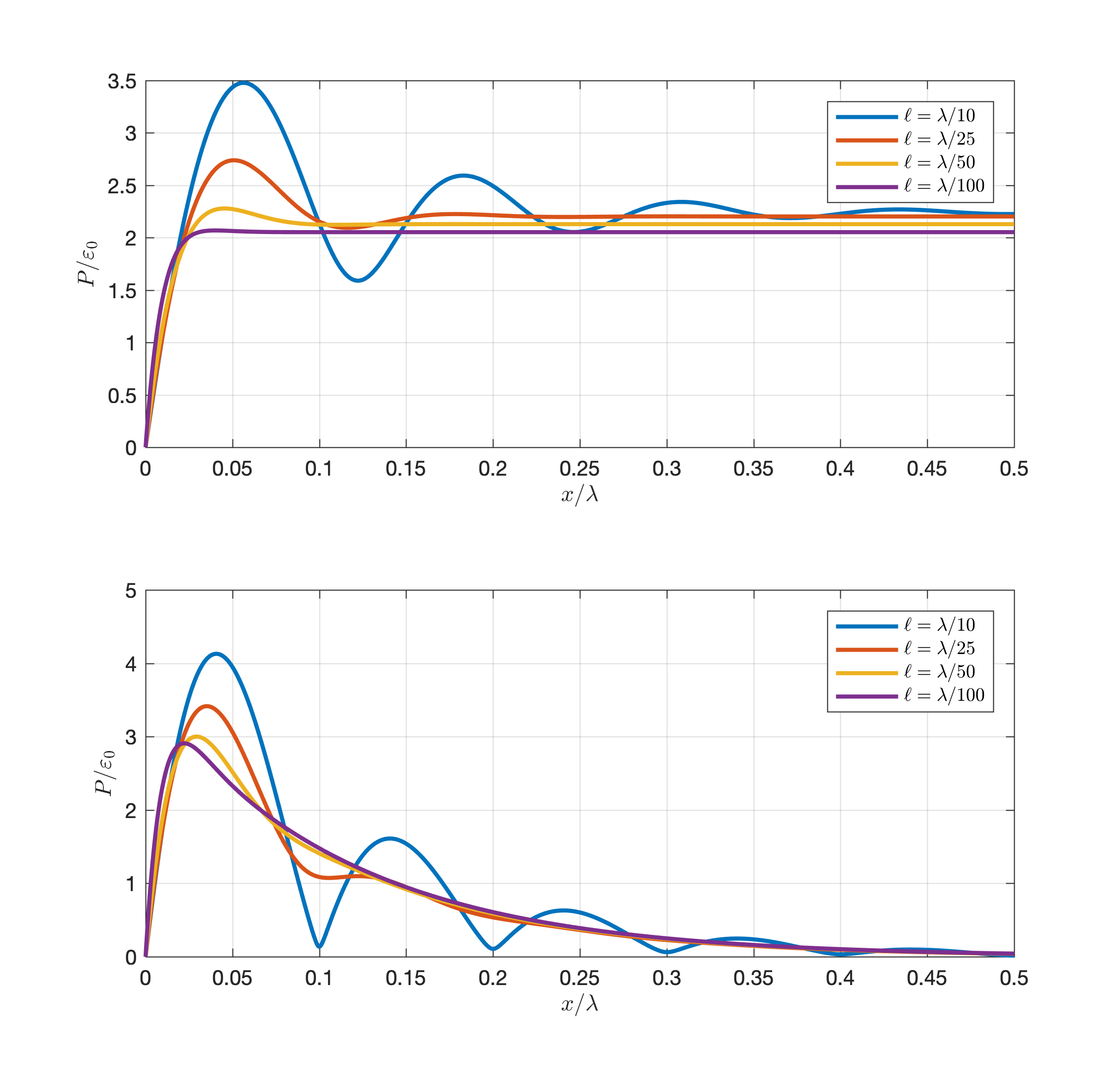}
	\caption{Polarization vector in a vacuum-solid interface when the solid is a non-local dielectric (upper panel) or conductor (lower panel). We see how, although the polarization cancels at the surface, as we reduce the characteristic length the polarization tends towards a step discontinuity in the case of the non-local dielectric and to concentrate all the current at the surface for the conductor, as predicted by local eletrdodynamics but also in agreement with the present theory. }
	\label{fig:Polarization}
\end{figure}

Finally, figure \ref{fig:Polarization} shows the behaviour of the polarization vector at a vacuum-solid interface, when the solid has a dielectric (upper panel) or a conductor (lower panel) character, with parameters $\beta_b=2$, $\chi_{Eb}=\pm 3$, with the +(-) sign corresponding to the dielectric (conductor). For the conductor case we also set $k_S=10k_0$ and results are shown for $\ell_b=\lambda/10,\lambda/25,\lambda/50$ and $\lambda/100$. Boundary conditions imply the cancelation of the polarization at the interface, but we see how, as the characteristic length becomes smaller, we recover the local behaviour for both types of materials, and the polarization presents a step discontinuity for the dielectric while it is concentrated at the surface for the conductor. It has to be pointed out that, within the hydrodynamic description, the behaviour of the conductor could not be described in this way, since at normal incidence it is identical to a local metal, so that this description is clearly more correct if microscopic arguments are to be included.

%%%%%%%%%%%%%%%%%%%%%%%%%%%%%%%%%%%%%%%%%%%%%%%%%%%%%%%%%%%%%%%%%%%%%%%%%%%%%%%
%%%%%%%%%%%%%%%%%%%%%%%%%%%%%%%%%%%%%%%%%%%%%%%%%%%%%%%%%%%%%%%%%%%%%%%%%%%%%%%
%%%%%%%%%%%%%%%%%%%%%%%%%%%%%%%%%%%%%%%%%%%%%%%%%%%%%%%%%%%%%%%%%%%%%%%%%%%%%%%
%%%%%%%%%%%%%%%%%%%%%%%%%%%%%%%%%%%%%%%%%%%%%%%%%%%%%%%%%%%%%%%%%%%%%%%%%%%%%%%

%%%%%%%%%%%%%%%%%%%%%%%%%%%%%%%%%%%%%%%%%%%%%%%%%%%%%%%%%%%%%%%%%%%%%%%%%%%%%%%

The presented approach is clearly consistent in the limiting situations, in which we recover the ``traditional'' boundary conditions as a progressive limiting situation, something that is not given in other approaches, which just define a set of boundary conditions but do not specify when these are valid or not. 

%%%%%%%%%%%%%%%%%%%%%%%%%%%%%%%%%%%%%%%%%%%%%%%%%%%%%%%%%%%%%%%%%%%%%%%%%%%%%

%%%%%%%%%%%%%%%%%%%%%%%%%%%%%%%%%%%%%%%%%%%%%%%%%%%%%%%%%%%%%%%%%%%%%%%%%%%%%%%%%%%%%%%%%%%
%%%%%%%%%%%%%%%%%%%%%%%%%%%%%%%%%%%%%%%%%%%%%%%%%%%%%%%%%%%%%%%%%%%%%%%%%%%%%%%%%%%%%%%%%%%
%%%%%%%%%%%%%%%%%%%%%%%%%%%%%%%%%%%%%%%%%%%%%%%%%%%%%%%%%%%%%%%%%%%%%%%%%%%%%%%%%%%%%%%%%%%
%\subsection{Local Material Limit}
%
%
%\subeqs{
%\fE(\fr,\omega)&=(e^{ik_0y}+Re^{-ik_0y})\fx\\
%\fB(\fr,\omega)&=-\frac{k_0}{\omega}(e^{ik_0y}-Re^{-ik_0y})\fz
%}
%
%\subeqs{
%\fE(\fr,\omega)&=(T_+e^{ik_{+}y}+T_-e^{ik_{-}y})\fx\\
%\fB(\fr,\omega)&=-\frac{1}{\omega}(k_+T_+e^{ik_{+}y}+k_-T_-e^{ik_{-}y})\fz\\
%%\fJ(\fr,\omega)&=\left(\frac{T_+}{\gamma k_0^2-\beta k_+^2}e^{ik_{+}y}+\frac{T_-}{\gamma k_0^2-\beta k_-^2}e^{ik_{-}y}\right)\fx
%\fJ(\fr,\omega)&=i\omega\eps_0\left(\frac{k_0^2-k_+^2}{k_0^2}T_+e^{ik_{+}y}+\frac{k_0^2-k_-^2}{k_0^2}T_-e^{ik_{-}y}\right)\fx
%}
%
%\subeqs{
%1+R&=T_++T_-\\
%1-R&=\frac{k_+}{k_0}(T_++\frac{k_-}{k_+}T_-)\\
%0&=\left(\frac{k_0^2-k_+^2}{k_0^2}T_++\frac{k_0^2-k_-^2}{k_0^2}T_-\right)
%}
%
%
\subsubsection{Oblique incidence}
In the previous subsection we considered the different interfaces at normal incidence, and this analysis allowed us to study the behaviour of the fields when material's discontinuities appear. In most of the real problems the fields do not incide normally at a surface, as for example in cylindrical or spherical objects, therefore the analysis of the behaviour of the fields when a given angle $\theta_0$ is formed with the normal to the surface will help us to understand the influence of spatial dispersion in more general problems. As will be seen below, the main difference is that the longitudinal mode is excited when a normal component of the electric field appears at the interface. This longitudinal mode will be required to satisfy the continuity of the fields, and it will play a similar role as the evanescent mode studied before. 

For an isotropic material the possible solutions of the electromagnetic field can be decomposed in one longitudinal ($L$) and two transverse ($T$) modes, and each of the transverse modes has two components, which we define as the $S_1, S_2$ and $P_1,P_2$ polarizations. 

Since the component of the wavevector parallel to the interface is a conserved quantity and, consequently, identical to all the polarizations, we define  the wavevector as
\beq
\fK_{i\sigma}^{\pm}=\pm q_{i\sigma}\fn+\fK_t,
\eeq
where the index $i$ indicates in which material the wave propagates ($i=a,b$ in our case) and $\sigma=L,S_1, S_2, P_1, P_ 2$. The vector $\fK_t$ is the component of the wavevector parallel to the surface, therefore the component of the wavevector normal to the surface is given by 
\beq
q_{i\sigma}=\sqrt{k_{i\sigma}^2-|\fK_t|^2},
\eeq
with $k_{i\sigma}=|\fK_{i\sigma}^{\pm}|$ being $k_L$ for $\sigma=L$ and $k_T$ for the different transverse modes, i.e., the solutions of equations \eqref{eq:kL} and \eqref{eq:kT}. 

The unit vectors parallel to these polarizations are given by
\subeqs{
\fu_L&=\frac{\fK_L}{|\fK_L|},\\
\fu_S&=\fn\times\frac{\fK_T}{|\fK_T|},\\
\fu_P&=\fu_S\times\frac{\fK_T}{|\fK_T|},
}
and the following relationships are found
\subeqs{
\fK_L\times\fu_L&=0\\
\fK_T\times\fu_S&=k_T\fu_P\\
\fK_T\times\fu_P&=-k_T\fu_S.
}

It has to be pointed out that the $\fu_S$ and $\fu_P$ are orthogonal for the same root of equation \eqref{eq:kT}, but not for different roots. Also, $\fu_L$ is not in general orthogonal to these vectors.

We assume that material $a$ is vacuum, and that the incident electric field is a plane wave with a general polarization state, thus

\beq
\fE_0=\sum_{\sigma=S,P}A_\sigma e^{iq_{a\sigma}n}e^{i\fK_t\cdot\fr}\fu_\sigma^+
\eeq
which excites a reflected field given by 
\beq
\fE_R=\sum_{\sigma=S,P}B_\sigma e^{-iq_{a\sigma}n}e^{i\fK_t\cdot\fr}\fu_\sigma^-.
\eeq

Material $b$ is a non-local solid, therefore we will have two solutions for each of the $S$ and $P$ polarizations plus the longitudinal mode $L$, therefore the transmitted electric field will be
\beq
\fE_T=\sum_{\sigma=S_j,P_j,L}C_\sigma e^{iq_{b\sigma}n}e^{i\fK_t\cdot\fr}\fu_\sigma^+
\eeq
where $\sigma=S_1,S_2,P_1,P_2$ and $L$. Related with each component of the electric field we have the magnetic field
\subeqs{
\omega \fB_0&=\sum_{\sigma=S,P} A_\sigma e^{iq_{a\sigma}n}e^{i\fK_t\cdot\fr}\fK_{a\sigma}^+\times\fu_\sigma^+\\
\omega \fB_R&=\sum_{\sigma=S,P} B_\sigma e^{-iq_{a\sigma}n}e^{i\fK_t\cdot\fr}\fK_{a\sigma}^-\times\fu_\sigma^-\\
\omega \fB_T&=\sum_{\sigma=S_j,P_j,L} C_\sigma e^{iq_{b\sigma}n}e^{i\fK_t\cdot\fr}\fK_{b\sigma}^+\times\fu_\sigma^+
}
and the polarization vector
\beq
\begin{split}
-\frac{1}{\eps_0}\fP_T=\sum_{\sigma=S_j,P_j} \left(1-\frac{k_{b\sigma}^2}{k_0^2}\right)C_\sigma e^{iq_{b\sigma}n}e^{i\fK_t\cdot\fr}\fu_\sigma^++\\
C_Le^{iq_{b\sigma}n}e^{i\fK_t\cdot\fr}\fu_L^+
\end{split}
\eeq

The inputs of the system are the coefficients of the incident electric field $A_S$ and $A_P$, and we have to obtain the two reflected amplitudes $B_S$ and $B_P$ and the five transmitted amplitudes $C_{S1},C_{S2},C_{P1},C_{P2}$ and $C_L$. We need therefore, as discussed in section \ref{sec:BC}, seven equations, which correspond to the seven boundary conditions \eqref{eq:BCV1}-\eqref{eq:BCV2}.

It is easy to see that the $S$ polarization uncouples from the $P$ and $L$, which in turn means that the incident $A_S$ field excites only the $B_S$ reflected field and the $C_{S1}$ and $C_{S2}$ transmitted fields. The cancelation of the polarization vector parallel to $\fu_S$ implies
\beq
C_{S_2}=-\frac{k_0^2-k_{b1}^2}{k_0^2-k_{b2}^2}C_{S_1}
\eeq
while the continuity of the transverse components of the electric and magnetic fields gives
\subeqs{
A_S+B_S&=C_{S_1}+C_{S_2}\\
A_S-B_S&=\frac{q_{b1}}{q_{a0}}C_{S_1}+\frac{q_{b2}}{q_{a0}}C_{S_2}
}
from which we can obtain the reflection coefficient of the $S$ mode $R_S=B_S/A_S$ as
\beq
\frac{1-R_S}{1+R_S}=\frac{q_{b1}}{q_{a0}}\frac{(k_{b2}^2-k_0^2)-q_{b2}/q_{b1}(k_{b1}^2-k_0^2)}{k_{b2}^2-k_{b1}^2}.
\eeq

We can proceed similarly to obtain the reflection coefficient of the $P$ polarization $R_P=B_P/A_P$, using
\beq
\fu_P^+=-\frac{|\fK_t|^2}{k_{bj}^2}\fn+\frac{q_{bj}}{k_{bj}^2}\fK_t
\eeq
then the continuity of the transverse electric and magnetic field is 
\subeqs{
\frac{q_{a0}}{k_0^2}\left(A_P-B_P\right)&=\frac{q_{b1}}{k_{b1}^2}C_{P_1}+\frac{q_{b2}}{k_{b2}^2}C_{P_2}+\frac{1}{k_L}C_L\\
A_P+B_P&=C_{P_1}+C_{P_2}
}
while the cancelation of the polarization vector gives
\subeqs{
\frac{q_{bL}}{K_L}C_L-\frac{|\fK_t|^2}{k_{b1}^2}\left(1-\frac{k_{b1}^2}{k_0^2}\right)C_{P_1}-\frac{|\fK_t|^2}{k_{b2}^2}\left(1-\frac{k_{b2}^2}{k_0^2}\right)C_{P_2}&=0\\
\frac{1}{K_L}C_L+\frac{q_{b1}}{k_{b1}^2}\left(1-\frac{k_{b1}^2}{k_0^2}\right)C_{P_1}+\frac{q_{b2}}{k_{b2}^2}\left(1-\frac{k_{b2}^2}{k_0^2}\right)C_{P_2}&=0.
}

From the above two equations we can obtain the relationship of the $C_{P_1}$ with the $C_{P_2}$ and $C_L$ coefficients, giving
\subeqs{
C_{P_2}&=-\frac{k_{b2}^2}{k_{b1}^2}\frac{q_{bL}q_{b1}+|\fK_t|^2}{q_{bL}q_{b2}+|\fK_t|^2}\frac{k_0^2-k_{b1}^2}{k_0^2-k_{b2}^2}C_{P_1}\equiv \chi_P C_{P_1}\\
C_L&=k_L\frac{q_{b2}-q_{b1}}{q_{bL}q_{b2}-|\fK_t|^2}\frac{|\fK_t|^2(k_0^2-k_{b1}^2)}{k_0^2k_{b1}^2}C_{P1}\equiv \chi_L C_{P_1}
%\frac{q_{aL}}{K_L}C_L&=\frac{|\fK_t|^2}{k_{a1}^2}\left(1-\frac{k_{a1}^2}{k_0^2}\right)C_{P1}+\frac{|\fK_t|^2}{k_{a2}^2}\left(1-\frac{k_{a2}^2}{k_0^2}\right)C_{P2}
}
so that finally we obtain the $R_P$ coefficient as
\beq
\frac{1-R_P}{1+R_P}=\frac{k_0^2}{q_{a0}}\frac{q_{b1}/k_{b1}^2+q_{b2}/k_{b2}^2\chi_{P}+1/k_L\chi_L}{1+\chi_{P}}.
\eeq

At the interface between vacuum and a local dielectric material the normal component of the electric field is not continuous, and a bounded surface charge exists. This is due to the discontinuity of the polarizability of the material which induces a discontinuity in the polarization vector and, from the continuity equation,
\beq
\rho=-\nabla\cdot\fP=-i\fK\cdot\fP-\partial_nP_n\sim \delta(n)
\eeq
this discontinuity implies that it is actually the normal component of dielectric displacement $\fD=(1+\chi)\fE$ the quantity continuous through the interface. In the elastodynamic description however, due to the continuity of the polarization vector $\fP$, there are no bounded surface charges, consequently the normal component of the electric field is also continuous at an interface. 

 It is interesting to remark, as was explained before, that the shear modulus in the elastodynamic description is the quantity responsible of the non-local response of the material at normal incidence or at oblique but with the $S$ polarization. If we ignore this term, as it is the case in the hydrodynamic model, the non-local response is not observed and the material is indistinguishable from a local one. This is obviously an incoherence of the hydrodynamic description, since non-locality is a bulk property and should be observed independently of the polarizaiton of the field. 
%%%%%%%%%%%%%%%%%%%%%%%%%%%%%%%%%%%%%%%%%%%%%%%%%%%%%%%%%%%%%%%%%%%%%%%%%%%%%%%%%%%%%%%%%%
%%%%%%%%%%%%%%%%%%%%%%%%%%%%%%%%%%%%%%%%%%%%%%%%%%%%%%%%%%%%%%%%%%%%%%%%%%%%%%%%%%%%%%%%%%
\subsection{Scattering by a Cylinder}
Let us consider now the scattering of electromagnetic waves by a non-local cylinder. It is assumed that the axis of the cylinder is parallel to the $z$-axis, and that the wavevector lies on the $xy$-plane, so that there is no $z$ dependence on the fields. Under these conditions, the fields outside the cylinder are a combination of $E$ and $M$ polarizations, as usual. Then, the incident field to the cylinder is expressed as
\beq
\fE_0(\fr)=\sum_{q}A_{q}^{M}\nabla_t\times \hat{\bm{z}}J_q(k_0r)e^{iq\theta}+\sum_{q}A_{q}^{E} \hat{\bm{z}}J_q(k_0r)e^{iq\theta}
\eeq
while the scattered field as set as
\beq
\fE_{SC}(\fr)=\sum_{q}B_{q}^{M}\nabla_t\times \hat{\bm{z}}H_q(k_0r)e^{iq\theta}+\sum_{q}B_{q}^{E} \hat{\bm{z}}H_q(k_0r)e^{iq\theta}
\eeq

Inside te cylinder we have now the longitudinal mode $L$ plus the two transverse modes $k_{an}$, for $n=1,2$, which are decomposed as a $E$ and $M$ polarizations, as in vaccum. We have therefore that inside the cylinder the field is given by
\beq
\begin{split}
\fE_i(\fr)=\sum_{q}C_q^L\nabla_tJ_q(k_Lr)e^{iq\theta}+\\
\sum_{q,n}C_{qn}^{M}\nabla_t\times \hat{\bm{z}}J_q(k_{an}r)e^{iq\theta}+
\sum_{q,n}C_{qn}^{E} \hat{\bm{z}}J_q(k_{an}r)e^{iq\theta}
\end{split}
\eeq

The polar symmetry of the cylinder will uncouple all the harmonics labeled by the angular index $q$. Boundary conditions implies the continuity of the transverse components of the electric and magnetic fields plus the cancelation of the polarization at the boundary of the cylinder. Let us consider first the cancelation of the polarization at the boundary, since this field is defined only inside the cylinder, then we have
\beq
\label{eq:pol}
-\frac{1}{\eps_0}\fP=\sum_{\sigma=S,P,L}\left(1-\frac{k_{\sigma}^2}{k_0^2}\right)\fE_\sigma+\frac{1}{k_0^2}\nabla\nabla\cdot\fE
\eeq

The polarization vector is therefore parallel to the electric field plus the $\nabla\nabla\cdot\fE$ term, which is a contribution due to the longitudinal mode. It is obvious then that, since the field is independent of the $z$ coordinate, we will not have a coupling between the $L$ and $E$ polarizations, so that the cancelation of the $z$ component of the polarization vector will imply
\beq
C_{q2}^{E}=-\frac{k_0^2-k_{a1}^2}{k_0^2-k_{a2}^2}\frac{J_q(k_{a1}R_0)}{J_q(k_{a2}R_0)}C_{q1}^{E}\equiv Z_q^{21}C_{q1}^{E}
\eeq
while the usual continuity of the transverse components of the electric and magnetic field are
\begin{widetext}
\subeqs{
A_q^EJ_q(k_0R_0)+B_q^EH_q(k_0R_0)&=C_{q1}^EJ_q(k_{a1}R_0)+C_{q2}^EJ_q(k_{a2}R_0)\\
A_q^Ek_0J'_q(k_0R_0)+B_q^Ek_0H'_q(k_0R_0)&=C_{q1}^Ek_{a1}J'_q(k_{a1}R_0)+C_{q2}^Ek_{a2}J'_q(k_{a2}R_0)
}
\end{widetext}
The scattering properties of the cylinder are defined by means of the $T$-Matrix, which is defined as 
\beq
B_q^E=T_q^EA_q^E
\eeq
it is obvious then that
\beq
T_q^E=-\frac{J_q(k_0R_0)-Z_q^EJ'_q(k_0R_0)}{H_q(k_0R_0)-Z_q^EH'_q(k_0R_0}
\eeq
with
\beq
\label{eq:ZE}
Z_q^E=\frac{J_q(k_{a1}R_0)+Z_q^{21}J_q(k_{a2}R_0)}{k_{a1}J'_q(k_{a1}R_0)+Z_q^{21}k_{a2}J'_q(k_{a2}R_0)}k_0
\eeq

We see therefore how the scattering properties of the $E$ polarized wave are affected by the non-local response of the material, effect that is not considered by the hydrodynamic model. It seems now obvious that a large number of resonances will appear due to the complex frequency dependence of the impedance $Z_q^E$ defined by equation \eqref{eq:ZE}. 

Let us consider now the quasi-dielectric limit, in which the solutions for the wavenumbers $k_{a1}$ and $k_{a2}$ are given by equations \eqref{eq:ka1D} and \eqref{eq:ka2D}. We have therefore that while $k_{a1}$ is a propagative mode with a wavelength similar to that of the background, $k_{a2}$ is an evanescent mode with a tail length defined by $\ell_0$. Using $J_q(ix)=i^qI_q(x)$, with $I_q$ being the modified Bessel function of first class, we have, after some little algebra, that (removing the common factor $i^q$)
\beq
 Z_q^{21}=-\frac{k_0^2\ell_0^2-k_{a1}^2\ell_0^2}{k_0^2\ell_0^2+1}\frac{J_q(k_{a1}R_0)}{I_q(R_0/\ell_0)}
\eeq
and
\beq
\label{eq:ZE}
Z_q^E=\frac{J_q(k_{a1}R_0)+Z_q^{21}I_q(R_0/\ell_0)}{k_{a1}\ell_0J'_q(k_{a1}R_0)+Z_q^{21}I'_q(R_0/\ell_0)}k_0\ell_0
\eeq

Similarly, for the $M$ polarization, the cancelation of the polarization at the surface implies a coupling of the $M$ and $L$ modes inside the cylinder, with the following condition between coefficients,
\begin{widetext}
\subeqs{
C_{q}^Lk_LJ'_q(k_LR_0)+\frac{iq}{R_0}\left(1-\frac{k_{a1}^2}{k_0^2}\right)C_{q1}^MJ_q(k_{a1}R_0)+\frac{iq}{R_0}\left(1-\frac{k_{a2}^2}{k_0^2}\right)C_{q2}^MJ_q(k_{a2}R_0)&=0\\
-\frac{iq}{R_0}C_{q}^LJ_q(k_LR_0)+k_{a1}\left(1-\frac{k_{a1}^2}{k_0^2}\right)C_{q1}^MJ'_q(k_{a1}R_0)+k_{a2}\left(1-\frac{k_{a2}^2}{k_0^2}\right)C_{q2}^MJ'_q(k_{a2}R_0)&=0
}
\end{widetext}
from which we can solve for the $C_q^L$ and $C_{q2}^M$ coefficients as a function of the $C_{q1}^M$, solving then the scattering problem from the electromagnetic boundary conditions. The scattering coefficients now have a more complex expression now, as for the planar interface at oblique incidence. The expected number of modes is now richer than for the local case, as well as for the hydrodynamic model, which considers non-local effects only for the $M$ polarization but even in this case considers only the excitation of the $L$ mode. 

\subsection{Scattering by a Sphere}
Finally, let us consider the scattering by a non-local spherical object of radius $R_0$. Outside the sphere the fields are decomposed in the usual way\cite{morse1954methods},
\subeqs{
\fE(\fr)&=\bm{L}\psi^E(\fr)+\frac{i}{k_{0}}\nabla\times\bm{L}\psi^M(\fr)\\
i\omega\fB(\fr)&=\nabla\times\bm{L}\psi^E(\fr)+ik_0\bm{L}\psi^M(\fr)\\
}
with
\beq
\bm{L}=-i\fr\times\nabla
\eeq
and where $\psi^E(\fr)$ and $\psi^M(\fr)$ are the electric and magnetic potentials which can be expanded as
\subeqs{
\psi^E(\fr)&=\sum_{n,m} \left(A_{nm}^Ej_n(k_0r)+B_{nm}^Eh_n(k_0r)\right)Y_{nm}(\hat{\fr})\\
\psi^M(\fr)&=\sum_{n,m} \left(A_{nm}^Mj_n(k_0r)+B_{nm}^Mh_n(k_0r)\right)Y_{nm}(\hat{\fr})
}
Inside the sphere, we expand the modes in a similar way but including the longitudinal mode,
\beq
\fE(\fr)=\frac{1}{k_L}\nabla\psi^L(\fr)+\sum_{\alpha=1,2}\bm{L}\psi^E_\alpha(\fr)+\sum_{\alpha=1,2}\frac{i}{k_{T\alpha}}\nabla\times\bm{L}\psi^M_\alpha(\fr)
\eeq
where now
\subeqs{
\psi^L(\fr)&=\sum_{n,m} C_{nm}^Lj_n(k_Lr)Y_{nm}(\hat{\fr})\\
\psi^E_\alpha(\fr)&=\sum_{n,m} C_{nm}^{E\alpha}j_n(k_{T\alpha}r)Y_{nm}(\hat{\fr})\\
\psi^M_\alpha(\fr)&=\sum_{n,m} C_{nm}^{M\alpha}j_n(k_{T\alpha}r)Y_{nm}(\hat{\fr})
}

The magnetic field has a similar expression since the longitudinal mode does not contribute to it, thus
\beq
i\omega\fB(\fr)=\sum_{\alpha=1,2}\nabla\times\bm{L}\psi^E_\alpha(\fr)+\sum_{\alpha=1,2}ik_{T\alpha}\bm{L}\psi^M_\alpha(\fr)
\eeq
And the polarization is, using equation \eqref{eq:pol},
\begin{widetext}
\beq
-\frac{1}{\eps_0}\fP=\sum_{\alpha=1,2}\left(1-\frac{k_{T\alpha}^2}{k_0^2}\right)\bm{L}\psi^E_\alpha(\fr)+\sum_{\alpha=1,2}\frac{i}{k_{T\alpha}}\left(1-\frac{k_{T\alpha}^2}{k_0^2}\right)\nabla\times\bm{L}\psi^M_\alpha(\fr)+\frac{1}{k_L}\nabla\psi^L(\fr)
\eeq
\end{widetext}
The cancelation of all the components of the polarization implies the decoupling of the $E$ polarization, since the operator $\bm{L}$ is orthogonal to both the $\nabla$ and the $\nabla\times\bm{L}$ ones, thus we have
\beq
C_{nm}^{E2}=-\frac{k_0^2-k_{a1}^2}{k_0^2-k_{a2}^2}\frac{j_n(k_{a1}R_0)}{j_n(k_{a2}R_0)}C_{nm}^{E1}\equiv Z_{nm}^{21}C_{nm}^{E1}
\eeq
The cancelation of the components parallel to $\fr$ and to $\fr\times\bm{L}$ couple the $L$ and $M$ polarizations, and these are\cite{morse1954methods}
\begin{widetext}
\subeqs{
j'_n(k_LR_0)C_{nm}^L-n(n+1)\sum_{\alpha=1,2}\frac{j_n(k_{T\alpha}R_0)}{k_{T\alpha}R_0}\left(1-\frac{k_{T\alpha}^2}{k_0^2}\right)C_{nm}^{M\alpha}&=0\\
-\frac{j_n(k_LR_0)}{k_LR_0}C_{nm}^L+\sum_{\alpha=1,2}\left(\frac{j_n(k_{T\alpha}R_0)}{k_{T\alpha }R_0}+j'_n(k_{T\alpha}R_0)\right)\left(1-\frac{k_{T\alpha}^2}{k_0^2}\right)C_{nm}^{M\alpha}&=0
}
\end{widetext}
Solving from the above equations we find that 
\subeqs{
C_{nm}^L&=Z_{nm}^{L1}C_{nm}^{M1}\\
C_{nm}^{M2}&=Z_{nm}^{21}C_{nm}^{M1}
}

The continuity of the transverse components of the $\fE$ and $\fB$ fields define the scattering properties of the sphere by means of the $T_{n}^E$ and $T_n^M$ matrices, defined as
\subeqs{
B_{nm}^E&=T_n^EA_{nm}^E\\
B_{nm}^M&=T_n^MA_{nm}^E
}
since there is no coupling between modes. Finally, it is easy to see that 
\subeqs{
T_n^E&=-\frac{j_n(k_0R_0)-Z_n^E[k_0R_0j_n(k_0R_0]'}{h_n(k_0R_0)-Z_n^E[k_0R_0h_n(k_0R_0]'}\\
T_n^M&=-\frac{j_n(k_0R_0)-Z_n^M[k_0R_0j_n(k_0R_0]'}{h_n(k_0R_0)-Z_n^M[k_0R_0h_n(k_0R_0]'}.
}
\color{black}

%%%%%%%%%%%%%%%%%%%%%%%%%%%%%%%%%%%%%%%%%%%%%%%%%%%%%%%%%%%%%%%%%%%%%%%%%%%%%%%%%%%%%%%%%%%
%%%%%%%%%%%%%%%%%%%

%%%%%%%%%%%%%%%%%%%%%%%%%%%%%%%%%%%%%%%%%%%%%%%%%%%%%%%%%%%%%%%%%%%%%%%%
\section{Summary}
\label{sec:SM}
In summary, a self-consistent theory for the interaction of the classical electromagnetic field and matter has been presented. The theory, developed within the framework of elastodynamics, is based on a polarization vector whose equation of motion is identical to that of the elastodynamic field, and a set of boundary conditions arise in a natural way whose number is consistent with the number of electrodynamic modes. Elementary considerations about the limiting value of the parameters of the model allow us to define vacuum, local dielectrics, real conductors and hydrodynamic plasmas, recovering in each situation the boundary conditions employed in the literature. This description however includes the possibility of more advanced interfaces, composites and symmetries, since a full anisotropic description has been considered. 

This approach contains as well the basis  for the consideration of more advanced phenomena, like the spill-out of electrons across interfaces. We have proposed two methods to work on this issue, by considering vacuum a low-density material with some homogeneous or inhomogeneous electron density and by developing a properly defined density or current functional theories. Both approaches fits perfectly well within the present description, although further development of these is required.

Therefore, the theory presented here can be a starting point for any model of matter at the nanoscale, where the effects of spatial dispersion and continuity of the microscopic fields can be more relevant, and more refined models of matter are required.

This unified theory for the description of the interaction of light and matter at the nanoscale will doubtless provide a new and generalized description of nanophotonic structures which will be fundamental for either the characterization of nanomaterials and the accurate design of new nanophotonic devices.

%%%%%%%%%%%%%%%%%%%%%%%%%%%%%%%%%%%%%%%%%%%%%%%%%%%%%%%%%%%%%%%%%%%%%%%%%%%%%%%%%%%%%%%%%%%
%%%%%%%%%%%%%%%%%%%%%%%%%%%%%%%%%%%%%%%%%%%%%%%%%%%%%%%%%%%%%%%%%%%%%%%%%%%%%%%%%%%%%%%%%%%
\begin{acknowledgments}
Daniel Torrent acknowledges financial support through the ``Ram\'on y Cajal'' fellowship under grant number RYC-2016-21188 and to the Ministry of Science, Innovation and Universities through Project No. RTI2018- 093921-A-C42.  JV Alvarez aknowledges financial support to the Ministry of Science, Innovation and Universities through Projects FIS2015-64886-C5-5-P and PGC2018-096955-B-C42.
\end{acknowledgments}
%\appendix
%\section{Vector Spherical Waves}
%\begin{widetext}
%\subeqs{
%\frac{1}{k_L}\nabla j_n(k_{L}r)Y_{n}^m(\theta,\phi)&=j'_n(k_Lr)Y_{n}^m(\theta,\phi)\hat{\fr}-i\frac{j_n(k_Lr)}{k_Lr}\hat{\fr}\times\bm{L}Y_{n}^m(\theta,\phi)\\
%\frac{i}{k_{T\alpha}}\nabla\times\bm{L} j_n(k_{T\alpha }r)Y_{n}^m(\theta,\phi)&=-n(n+1)\frac{j_n(k_{T\alpha}r)}{k_{T\alpha}r}Y_{n}^m(\theta,\phi)\hat{\fr}+
%i\left(\frac{j_n(k_{T\alpha}r)}{k_{T\alpha }r}+j'_n(k_{T\alpha}r)\right)\hat{\fr}\times\bm{L}Y_{n}^m(\theta,\phi)
%}
%\end{widetext}

%\bibliographystyle{apsrev}
%\bibliography{bibliography}

\begin{thebibliography}{52}
\expandafter\ifx\csname natexlab\endcsname\relax\def\natexlab#1{#1}\fi
\expandafter\ifx\csname bibnamefont\endcsname\relax
  \def\bibnamefont#1{#1}\fi
\expandafter\ifx\csname bibfnamefont\endcsname\relax
  \def\bibfnamefont#1{#1}\fi
\expandafter\ifx\csname citenamefont\endcsname\relax
  \def\citenamefont#1{#1}\fi
\expandafter\ifx\csname url\endcsname\relax
  \def\url#1{\texttt{#1}}\fi
\expandafter\ifx\csname urlprefix\endcsname\relax\def\urlprefix{URL }\fi
\providecommand{\bibinfo}[2]{#2}
\providecommand{\eprint}[2][]{\url{#2}}

\bibitem[{\citenamefont{Kelly et~al.}(2003)\citenamefont{Kelly, Coronado, Zhao,
  and Schatz}}]{kelly2003optical}
\bibinfo{author}{\bibfnamefont{K.~L.} \bibnamefont{Kelly}},
  \bibinfo{author}{\bibfnamefont{E.}~\bibnamefont{Coronado}},
  \bibinfo{author}{\bibfnamefont{L.~L.} \bibnamefont{Zhao}}, \bibnamefont{and}
  \bibinfo{author}{\bibfnamefont{G.~C.} \bibnamefont{Schatz}},
  \emph{\bibinfo{title}{The optical properties of metal nanoparticles: the
  influence of size, shape, and dielectric environment}}
  (\bibinfo{year}{2003}).

\bibitem[{\citenamefont{Amendola et~al.}(2017)\citenamefont{Amendola, Pilot,
  Frasconi, Marago, and Iati}}]{amendola2017surface}
\bibinfo{author}{\bibfnamefont{V.}~\bibnamefont{Amendola}},
  \bibinfo{author}{\bibfnamefont{R.}~\bibnamefont{Pilot}},
  \bibinfo{author}{\bibfnamefont{M.}~\bibnamefont{Frasconi}},
  \bibinfo{author}{\bibfnamefont{O.~M.} \bibnamefont{Marago}},
  \bibnamefont{and} \bibinfo{author}{\bibfnamefont{M.~A.} \bibnamefont{Iati}},
  \bibinfo{journal}{Journal of Physics: Condensed Matter}
  \textbf{\bibinfo{volume}{29}}, \bibinfo{pages}{203002}
  (\bibinfo{year}{2017}).

\bibitem[{\citenamefont{Burda et~al.}(2005)\citenamefont{Burda, Chen,
  Narayanan, and El-Sayed}}]{burda2005chemistry}
\bibinfo{author}{\bibfnamefont{C.}~\bibnamefont{Burda}},
  \bibinfo{author}{\bibfnamefont{X.}~\bibnamefont{Chen}},
  \bibinfo{author}{\bibfnamefont{R.}~\bibnamefont{Narayanan}},
  \bibnamefont{and} \bibinfo{author}{\bibfnamefont{M.~A.}
  \bibnamefont{El-Sayed}}, \bibinfo{journal}{Chemical reviews}
  \textbf{\bibinfo{volume}{105}}, \bibinfo{pages}{1025} (\bibinfo{year}{2005}).

\bibitem[{\citenamefont{Matricardi et~al.}(2018)\citenamefont{Matricardi,
  Hanske, Garcia-Pomar, Langer, Mihi, and Liz-Marzan}}]{matricardi2018gold}
\bibinfo{author}{\bibfnamefont{C.}~\bibnamefont{Matricardi}},
  \bibinfo{author}{\bibfnamefont{C.}~\bibnamefont{Hanske}},
  \bibinfo{author}{\bibfnamefont{J.~L.} \bibnamefont{Garcia-Pomar}},
  \bibinfo{author}{\bibfnamefont{J.}~\bibnamefont{Langer}},
  \bibinfo{author}{\bibfnamefont{A.}~\bibnamefont{Mihi}}, \bibnamefont{and}
  \bibinfo{author}{\bibfnamefont{L.~M.} \bibnamefont{Liz-Marzan}},
  \bibinfo{journal}{ACS nano} \textbf{\bibinfo{volume}{12}},
  \bibinfo{pages}{8531} (\bibinfo{year}{2018}).

\bibitem[{\citenamefont{Espinha et~al.}(2018)\citenamefont{Espinha, Dore,
  Matricardi, Alonso, Go{\~n}i, and Mihi}}]{espinha2018hydroxypropyl}
\bibinfo{author}{\bibfnamefont{A.}~\bibnamefont{Espinha}},
  \bibinfo{author}{\bibfnamefont{C.}~\bibnamefont{Dore}},
  \bibinfo{author}{\bibfnamefont{C.}~\bibnamefont{Matricardi}},
  \bibinfo{author}{\bibfnamefont{M.~I.} \bibnamefont{Alonso}},
  \bibinfo{author}{\bibfnamefont{A.~R.} \bibnamefont{Go{\~n}i}},
  \bibnamefont{and} \bibinfo{author}{\bibfnamefont{A.}~\bibnamefont{Mihi}},
  \bibinfo{journal}{Nature photonics} \textbf{\bibinfo{volume}{12}},
  \bibinfo{pages}{343} (\bibinfo{year}{2018}).

\bibitem[{\citenamefont{Raza et~al.}(2015)\citenamefont{Raza, Bozhevolnyi,
  Wubs, and Mortensen}}]{raza2015nonlocal}
\bibinfo{author}{\bibfnamefont{S.}~\bibnamefont{Raza}},
  \bibinfo{author}{\bibfnamefont{S.~I.} \bibnamefont{Bozhevolnyi}},
  \bibinfo{author}{\bibfnamefont{M.}~\bibnamefont{Wubs}}, \bibnamefont{and}
  \bibinfo{author}{\bibfnamefont{N.~A.} \bibnamefont{Mortensen}},
  \bibinfo{journal}{Journal of Physics: Condensed Matter}
  \textbf{\bibinfo{volume}{27}}, \bibinfo{pages}{183204}
  (\bibinfo{year}{2015}).

\bibitem[{\citenamefont{Aizpurua et~al.}(2003)\citenamefont{Aizpurua, Hanarp,
  Sutherland, K{\"a}ll, Bryant, and De~Abajo}}]{aizpurua2003optical}
\bibinfo{author}{\bibfnamefont{J.}~\bibnamefont{Aizpurua}},
  \bibinfo{author}{\bibfnamefont{P.}~\bibnamefont{Hanarp}},
  \bibinfo{author}{\bibfnamefont{D.}~\bibnamefont{Sutherland}},
  \bibinfo{author}{\bibfnamefont{M.}~\bibnamefont{K{\"a}ll}},
  \bibinfo{author}{\bibfnamefont{G.~W.} \bibnamefont{Bryant}},
  \bibnamefont{and} \bibinfo{author}{\bibfnamefont{F.~G.}
  \bibnamefont{De~Abajo}}, \bibinfo{journal}{Physical review letters}
  \textbf{\bibinfo{volume}{90}}, \bibinfo{pages}{057401}
  (\bibinfo{year}{2003}).

\bibitem[{\citenamefont{Myroshnychenko
  et~al.}(2008)\citenamefont{Myroshnychenko, Rodr{\'\i}guez-Fern{\'a}ndez,
  Pastoriza-Santos, Funston, Novo, Mulvaney, Liz-Marz{\'a}n, and
  de~Abajo}}]{myroshnychenko2008modelling}
\bibinfo{author}{\bibfnamefont{V.}~\bibnamefont{Myroshnychenko}},
  \bibinfo{author}{\bibfnamefont{J.}~\bibnamefont{Rodr{\'\i}guez-Fern{\'a}ndez}},
  \bibinfo{author}{\bibfnamefont{I.}~\bibnamefont{Pastoriza-Santos}},
  \bibinfo{author}{\bibfnamefont{A.~M.} \bibnamefont{Funston}},
  \bibinfo{author}{\bibfnamefont{C.}~\bibnamefont{Novo}},
  \bibinfo{author}{\bibfnamefont{P.}~\bibnamefont{Mulvaney}},
  \bibinfo{author}{\bibfnamefont{L.~M.} \bibnamefont{Liz-Marz{\'a}n}},
  \bibnamefont{and} \bibinfo{author}{\bibfnamefont{F.~J.~G.}
  \bibnamefont{de~Abajo}}, \bibinfo{journal}{Chemical Society Reviews}
  \textbf{\bibinfo{volume}{37}}, \bibinfo{pages}{1792} (\bibinfo{year}{2008}).

\bibitem[{\citenamefont{Baumberg et~al.}(2019)\citenamefont{Baumberg, Aizpurua,
  Mikkelsen, and Smith}}]{baumberg2019extreme}
\bibinfo{author}{\bibfnamefont{J.~J.} \bibnamefont{Baumberg}},
  \bibinfo{author}{\bibfnamefont{J.}~\bibnamefont{Aizpurua}},
  \bibinfo{author}{\bibfnamefont{M.~H.} \bibnamefont{Mikkelsen}},
  \bibnamefont{and} \bibinfo{author}{\bibfnamefont{D.~R.} \bibnamefont{Smith}},
  \bibinfo{journal}{Nature materials} p.~\bibinfo{pages}{1}
  (\bibinfo{year}{2019}).

\bibitem[{\citenamefont{Ciraci et~al.}(2013)\citenamefont{Ciraci, Pendry, and
  Smith}}]{ciraci2013hydrodynamic}
\bibinfo{author}{\bibfnamefont{C.}~\bibnamefont{Ciraci}},
  \bibinfo{author}{\bibfnamefont{J.~B.} \bibnamefont{Pendry}},
  \bibnamefont{and} \bibinfo{author}{\bibfnamefont{D.~R.} \bibnamefont{Smith}},
  \bibinfo{journal}{ChemPhysChem} \textbf{\bibinfo{volume}{14}},
  \bibinfo{pages}{1109} (\bibinfo{year}{2013}).

\bibitem[{\citenamefont{Jackson}(1999)}]{jackson1999classical}
\bibinfo{author}{\bibfnamefont{J.~D.} \bibnamefont{Jackson}},
  \emph{\bibinfo{title}{Classical electrodynamics}} (\bibinfo{year}{1999}).

\bibitem[{\citenamefont{Adler}(1962)}]{adler1962quantum}
\bibinfo{author}{\bibfnamefont{S.~L.} \bibnamefont{Adler}},
  \bibinfo{journal}{Physical Review} \textbf{\bibinfo{volume}{126}},
  \bibinfo{pages}{413} (\bibinfo{year}{1962}).

\bibitem[{\citenamefont{Agranovich and Ginzburg}(2013)}]{agranovich2013crystal}
\bibinfo{author}{\bibfnamefont{V.~M.} \bibnamefont{Agranovich}}
  \bibnamefont{and} \bibinfo{author}{\bibfnamefont{V.}~\bibnamefont{Ginzburg}},
  \emph{\bibinfo{title}{Crystal optics with spatial dispersion, and excitons}},
  vol.~\bibinfo{volume}{42} (\bibinfo{publisher}{Springer Science \& Business
  Media}, \bibinfo{year}{2013}).

\bibitem[{\citenamefont{Agranovich and
  Gartstein}(2006)}]{agranovich2006spatial}
\bibinfo{author}{\bibfnamefont{V.~M.} \bibnamefont{Agranovich}}
  \bibnamefont{and} \bibinfo{author}{\bibfnamefont{Y.~N.}
  \bibnamefont{Gartstein}}, \bibinfo{journal}{Physics-Uspekhi}
  \textbf{\bibinfo{volume}{49}}, \bibinfo{pages}{1029} (\bibinfo{year}{2006}).

\bibitem[{\citenamefont{Garcia~de Abajo}(2008)}]{garcia2008nonlocal}
\bibinfo{author}{\bibfnamefont{F.~J.} \bibnamefont{Garcia~de Abajo}},
  \bibinfo{journal}{The Journal of Physical Chemistry C}
  \textbf{\bibinfo{volume}{112}}, \bibinfo{pages}{17983}
  (\bibinfo{year}{2008}).

\bibitem[{\citenamefont{Belov et~al.}(2003)\citenamefont{Belov, Marques,
  Maslovski, Nefedov, Silveirinha, Simovski, and Tretyakov}}]{belov2003strong}
\bibinfo{author}{\bibfnamefont{P.}~\bibnamefont{Belov}},
  \bibinfo{author}{\bibfnamefont{R.}~\bibnamefont{Marques}},
  \bibinfo{author}{\bibfnamefont{S.}~\bibnamefont{Maslovski}},
  \bibinfo{author}{\bibfnamefont{I.}~\bibnamefont{Nefedov}},
  \bibinfo{author}{\bibfnamefont{M.}~\bibnamefont{Silveirinha}},
  \bibinfo{author}{\bibfnamefont{C.}~\bibnamefont{Simovski}}, \bibnamefont{and}
  \bibinfo{author}{\bibfnamefont{S.}~\bibnamefont{Tretyakov}},
  \bibinfo{journal}{Physical Review B} \textbf{\bibinfo{volume}{67}},
  \bibinfo{pages}{113103} (\bibinfo{year}{2003}).

\bibitem[{\citenamefont{Menzel et~al.}(2010)\citenamefont{Menzel, Paul,
  Rockstuhl, Pertsch, Tretyakov, and Lederer}}]{menzel2010validity}
\bibinfo{author}{\bibfnamefont{C.}~\bibnamefont{Menzel}},
  \bibinfo{author}{\bibfnamefont{T.}~\bibnamefont{Paul}},
  \bibinfo{author}{\bibfnamefont{C.}~\bibnamefont{Rockstuhl}},
  \bibinfo{author}{\bibfnamefont{T.}~\bibnamefont{Pertsch}},
  \bibinfo{author}{\bibfnamefont{S.}~\bibnamefont{Tretyakov}},
  \bibnamefont{and} \bibinfo{author}{\bibfnamefont{F.}~\bibnamefont{Lederer}},
  \bibinfo{journal}{Physical Review B} \textbf{\bibinfo{volume}{81}},
  \bibinfo{pages}{035320} (\bibinfo{year}{2010}).

\bibitem[{\citenamefont{Al{\`u}}(2011)}]{alu2011restoring}
\bibinfo{author}{\bibfnamefont{A.}~\bibnamefont{Al{\`u}}},
  \bibinfo{journal}{Physical Review B} \textbf{\bibinfo{volume}{83}},
  \bibinfo{pages}{081102} (\bibinfo{year}{2011}).

\bibitem[{\citenamefont{Chipouline et~al.}(2012)\citenamefont{Chipouline,
  Simovski, and Tretyakov}}]{chipouline2012basics}
\bibinfo{author}{\bibfnamefont{A.}~\bibnamefont{Chipouline}},
  \bibinfo{author}{\bibfnamefont{C.}~\bibnamefont{Simovski}}, \bibnamefont{and}
  \bibinfo{author}{\bibfnamefont{S.}~\bibnamefont{Tretyakov}},
  \bibinfo{journal}{Metamaterials} \textbf{\bibinfo{volume}{6}},
  \bibinfo{pages}{77} (\bibinfo{year}{2012}).

\bibitem[{\citenamefont{Mnasri et~al.}(2018)\citenamefont{Mnasri,
  Khrabustovskyi, Stohrer, Plum, and Rockstuhl}}]{mnasri2018beyond}
\bibinfo{author}{\bibfnamefont{K.}~\bibnamefont{Mnasri}},
  \bibinfo{author}{\bibfnamefont{A.}~\bibnamefont{Khrabustovskyi}},
  \bibinfo{author}{\bibfnamefont{C.}~\bibnamefont{Stohrer}},
  \bibinfo{author}{\bibfnamefont{M.}~\bibnamefont{Plum}}, \bibnamefont{and}
  \bibinfo{author}{\bibfnamefont{C.}~\bibnamefont{Rockstuhl}},
  \bibinfo{journal}{Physical Review B} \textbf{\bibinfo{volume}{97}},
  \bibinfo{pages}{075439} (\bibinfo{year}{2018}).

\bibitem[{\citenamefont{Mnasri et~al.}(2019{\natexlab{a}})\citenamefont{Mnasri,
  Khrabustovskyi, Plum, and Rockstuhl}}]{mnasri2019retrieving}
\bibinfo{author}{\bibfnamefont{K.}~\bibnamefont{Mnasri}},
  \bibinfo{author}{\bibfnamefont{A.}~\bibnamefont{Khrabustovskyi}},
  \bibinfo{author}{\bibfnamefont{M.}~\bibnamefont{Plum}}, \bibnamefont{and}
  \bibinfo{author}{\bibfnamefont{C.}~\bibnamefont{Rockstuhl}},
  \bibinfo{journal}{Physical Review B} \textbf{\bibinfo{volume}{99}},
  \bibinfo{pages}{035442} (\bibinfo{year}{2019}{\natexlab{a}}).

\bibitem[{\citenamefont{Mnasri et~al.}(2019{\natexlab{b}})\citenamefont{Mnasri,
  Goffi, Plum, and Rockstuhl}}]{mnasri2019homogenization}
\bibinfo{author}{\bibfnamefont{K.}~\bibnamefont{Mnasri}},
  \bibinfo{author}{\bibfnamefont{F.~Z.} \bibnamefont{Goffi}},
  \bibinfo{author}{\bibfnamefont{M.}~\bibnamefont{Plum}}, \bibnamefont{and}
  \bibinfo{author}{\bibfnamefont{C.}~\bibnamefont{Rockstuhl}},
  \bibinfo{journal}{JOSA B} \textbf{\bibinfo{volume}{36}}, \bibinfo{pages}{F99}
  (\bibinfo{year}{2019}{\natexlab{b}}).

\bibitem[{\citenamefont{Rojas et~al.}(1988)\citenamefont{Rojas, Claro, and
  Fuchs}}]{rojas1988nonlocal}
\bibinfo{author}{\bibfnamefont{R.}~\bibnamefont{Rojas}},
  \bibinfo{author}{\bibfnamefont{F.}~\bibnamefont{Claro}}, \bibnamefont{and}
  \bibinfo{author}{\bibfnamefont{R.}~\bibnamefont{Fuchs}},
  \bibinfo{journal}{Physical Review B} \textbf{\bibinfo{volume}{37}},
  \bibinfo{pages}{6799} (\bibinfo{year}{1988}).

\bibitem[{\citenamefont{Pack et~al.}(2001)\citenamefont{Pack, Hietschold, and
  Wannemacher}}]{pack2001failure}
\bibinfo{author}{\bibfnamefont{A.}~\bibnamefont{Pack}},
  \bibinfo{author}{\bibfnamefont{M.}~\bibnamefont{Hietschold}},
  \bibnamefont{and}
  \bibinfo{author}{\bibfnamefont{R.}~\bibnamefont{Wannemacher}},
  \bibinfo{journal}{Optics communications} \textbf{\bibinfo{volume}{194}},
  \bibinfo{pages}{277} (\bibinfo{year}{2001}).

\bibitem[{\citenamefont{Silveirinha}(2006)}]{silveirinha2006additional}
\bibinfo{author}{\bibfnamefont{M.~G.} \bibnamefont{Silveirinha}},
  \bibinfo{journal}{IEEE transactions on antennas and propagation}
  \textbf{\bibinfo{volume}{54}}, \bibinfo{pages}{1766} (\bibinfo{year}{2006}).

\bibitem[{\citenamefont{McMahon et~al.}(2010)\citenamefont{McMahon, Gray, and
  Schatz}}]{mcmahon2010nonlocal}
\bibinfo{author}{\bibfnamefont{J.~M.} \bibnamefont{McMahon}},
  \bibinfo{author}{\bibfnamefont{S.~K.} \bibnamefont{Gray}}, \bibnamefont{and}
  \bibinfo{author}{\bibfnamefont{G.~C.} \bibnamefont{Schatz}},
  \bibinfo{journal}{The Journal of Physical Chemistry C}
  \textbf{\bibinfo{volume}{114}}, \bibinfo{pages}{15903}
  (\bibinfo{year}{2010}).

\bibitem[{\citenamefont{Maslovski et~al.}(2010)\citenamefont{Maslovski,
  Morgado, Silveirinha, Kaipa, and Yakovlev}}]{maslovski2010generalized}
\bibinfo{author}{\bibfnamefont{S.~I.} \bibnamefont{Maslovski}},
  \bibinfo{author}{\bibfnamefont{T.~A.} \bibnamefont{Morgado}},
  \bibinfo{author}{\bibfnamefont{M.~G.} \bibnamefont{Silveirinha}},
  \bibinfo{author}{\bibfnamefont{C.~S.} \bibnamefont{Kaipa}}, \bibnamefont{and}
  \bibinfo{author}{\bibfnamefont{A.~B.} \bibnamefont{Yakovlev}},
  \bibinfo{journal}{New Journal of Physics} \textbf{\bibinfo{volume}{12}},
  \bibinfo{pages}{113047} (\bibinfo{year}{2010}).

\bibitem[{\citenamefont{David and García~de Abajo}(2011)}]{david2011spatial}
\bibinfo{author}{\bibfnamefont{C.}~\bibnamefont{David}} \bibnamefont{and}
  \bibinfo{author}{\bibfnamefont{F.~J.} \bibnamefont{García~de Abajo}},
  \bibinfo{journal}{The Journal of Physical Chemistry C}
  \textbf{\bibinfo{volume}{115}}, \bibinfo{pages}{19470}
  (\bibinfo{year}{2011}).

\bibitem[{\citenamefont{Moreau et~al.}(2013)\citenamefont{Moreau, Ciraci, and
  Smith}}]{moreau2013impact}
\bibinfo{author}{\bibfnamefont{A.}~\bibnamefont{Moreau}},
  \bibinfo{author}{\bibfnamefont{C.}~\bibnamefont{Ciraci}}, \bibnamefont{and}
  \bibinfo{author}{\bibfnamefont{D.~R.} \bibnamefont{Smith}},
  \bibinfo{journal}{Physical Review B} \textbf{\bibinfo{volume}{87}},
  \bibinfo{pages}{045401} (\bibinfo{year}{2013}).

\bibitem[{\citenamefont{G{\'o}mez-Gra{\~n}a
  et~al.}(2016)\citenamefont{G{\'o}mez-Gra{\~n}a, Le~Beulze,
  Treguer-Delapierre, Mornet, Duguet, Grana, Cloutet, Hadziioannou, Leng,
  Salmon et~al.}}]{gomez2016hierarchical}
\bibinfo{author}{\bibfnamefont{S.}~\bibnamefont{G{\'o}mez-Gra{\~n}a}},
  \bibinfo{author}{\bibfnamefont{A.}~\bibnamefont{Le~Beulze}},
  \bibinfo{author}{\bibfnamefont{M.}~\bibnamefont{Treguer-Delapierre}},
  \bibinfo{author}{\bibfnamefont{S.}~\bibnamefont{Mornet}},
  \bibinfo{author}{\bibfnamefont{E.}~\bibnamefont{Duguet}},
  \bibinfo{author}{\bibfnamefont{E.}~\bibnamefont{Grana}},
  \bibinfo{author}{\bibfnamefont{E.}~\bibnamefont{Cloutet}},
  \bibinfo{author}{\bibfnamefont{G.}~\bibnamefont{Hadziioannou}},
  \bibinfo{author}{\bibfnamefont{J.}~\bibnamefont{Leng}},
  \bibinfo{author}{\bibfnamefont{J.-B.} \bibnamefont{Salmon}},
  \bibnamefont{et~al.}, \bibinfo{journal}{Materials Horizons}
  \textbf{\bibinfo{volume}{3}}, \bibinfo{pages}{596} (\bibinfo{year}{2016}).

\bibitem[{\citenamefont{Pekar}(1958)}]{pekar1958theory}
\bibinfo{author}{\bibfnamefont{S.}~\bibnamefont{Pekar}}, \bibinfo{journal}{Sov.
  Phys. JETP} \textbf{\bibinfo{volume}{6}}, \bibinfo{pages}{785}
  (\bibinfo{year}{1958}).

\bibitem[{\citenamefont{Melnyk and Harrison}(1970)}]{melnyk1970theory}
\bibinfo{author}{\bibfnamefont{A.~R.} \bibnamefont{Melnyk}} \bibnamefont{and}
  \bibinfo{author}{\bibfnamefont{M.~J.} \bibnamefont{Harrison}},
  \bibinfo{journal}{Physical Review B} \textbf{\bibinfo{volume}{2}},
  \bibinfo{pages}{835} (\bibinfo{year}{1970}).

\bibitem[{\citenamefont{Halevi and Fuchs}(1984)}]{halevi1984generalised}
\bibinfo{author}{\bibfnamefont{P.}~\bibnamefont{Halevi}} \bibnamefont{and}
  \bibinfo{author}{\bibfnamefont{R.}~\bibnamefont{Fuchs}},
  \bibinfo{journal}{Journal of Physics C: Solid State Physics}
  \textbf{\bibinfo{volume}{17}}, \bibinfo{pages}{3869} (\bibinfo{year}{1984}).

\bibitem[{\citenamefont{Yang et~al.}(2019)\citenamefont{Yang, Zhu, Yan,
  Agarwal, Zheng, Joannopoulos, Lalanne, Christensen, Berggren, and
  Solja{\v{c}}i{\'c}}}]{yang2019general}
\bibinfo{author}{\bibfnamefont{Y.}~\bibnamefont{Yang}},
  \bibinfo{author}{\bibfnamefont{D.}~\bibnamefont{Zhu}},
  \bibinfo{author}{\bibfnamefont{W.}~\bibnamefont{Yan}},
  \bibinfo{author}{\bibfnamefont{A.}~\bibnamefont{Agarwal}},
  \bibinfo{author}{\bibfnamefont{M.}~\bibnamefont{Zheng}},
  \bibinfo{author}{\bibfnamefont{J.~D.} \bibnamefont{Joannopoulos}},
  \bibinfo{author}{\bibfnamefont{P.}~\bibnamefont{Lalanne}},
  \bibinfo{author}{\bibfnamefont{T.}~\bibnamefont{Christensen}},
  \bibinfo{author}{\bibfnamefont{K.~K.} \bibnamefont{Berggren}},
  \bibnamefont{and}
  \bibinfo{author}{\bibfnamefont{M.}~\bibnamefont{Solja{\v{c}}i{\'c}}},
  \bibinfo{journal}{Nature} \textbf{\bibinfo{volume}{576}},
  \bibinfo{pages}{248} (\bibinfo{year}{2019}).

\bibitem[{\citenamefont{Henneberger}(1998)}]{henneberger1998additional}
\bibinfo{author}{\bibfnamefont{K.}~\bibnamefont{Henneberger}},
  \bibinfo{journal}{Physical review letters} \textbf{\bibinfo{volume}{80}},
  \bibinfo{pages}{2889} (\bibinfo{year}{1998}).

\bibitem[{\citenamefont{Luo et~al.}(2013)\citenamefont{Luo,
  Fernandez-Dominguez, Wiener, Maier, and Pendry}}]{luo2013surface}
\bibinfo{author}{\bibfnamefont{Y.}~\bibnamefont{Luo}},
  \bibinfo{author}{\bibfnamefont{A.}~\bibnamefont{Fernandez-Dominguez}},
  \bibinfo{author}{\bibfnamefont{A.}~\bibnamefont{Wiener}},
  \bibinfo{author}{\bibfnamefont{S.~A.} \bibnamefont{Maier}}, \bibnamefont{and}
  \bibinfo{author}{\bibfnamefont{J.}~\bibnamefont{Pendry}},
  \bibinfo{journal}{Physical review letters} \textbf{\bibinfo{volume}{111}},
  \bibinfo{pages}{093901} (\bibinfo{year}{2013}).

\bibitem[{\citenamefont{Kupresak et~al.}(2018)\citenamefont{Kupresak, Zheng,
  Vandenbosch, and Moshchalkov}}]{kupresak2018comparison}
\bibinfo{author}{\bibfnamefont{M.}~\bibnamefont{Kupresak}},
  \bibinfo{author}{\bibfnamefont{X.}~\bibnamefont{Zheng}},
  \bibinfo{author}{\bibfnamefont{G.~A.} \bibnamefont{Vandenbosch}},
  \bibnamefont{and} \bibinfo{author}{\bibfnamefont{V.~V.}
  \bibnamefont{Moshchalkov}}, \bibinfo{journal}{Advanced Theory and
  Simulations} \textbf{\bibinfo{volume}{1}}, \bibinfo{pages}{1800076}
  (\bibinfo{year}{2018}).

\bibitem[{\citenamefont{Royer and Dieulesaint}(1999)}]{royer1999elastic}
\bibinfo{author}{\bibfnamefont{D.}~\bibnamefont{Royer}} \bibnamefont{and}
  \bibinfo{author}{\bibfnamefont{E.}~\bibnamefont{Dieulesaint}},
  \emph{\bibinfo{title}{Elastic waves in solids I: Free and guided
  propagation}} (\bibinfo{publisher}{Springer Science \& Business Media},
  \bibinfo{year}{1999}).

\bibitem[{\citenamefont{Mortensen et~al.}(2014)\citenamefont{Mortensen, Raza,
  Wubs, S{\o}ndergaard, and Bozhevolnyi}}]{mortensen2014generalized}
\bibinfo{author}{\bibfnamefont{N.~A.} \bibnamefont{Mortensen}},
  \bibinfo{author}{\bibfnamefont{S.}~\bibnamefont{Raza}},
  \bibinfo{author}{\bibfnamefont{M.}~\bibnamefont{Wubs}},
  \bibinfo{author}{\bibfnamefont{T.}~\bibnamefont{S{\o}ndergaard}},
  \bibnamefont{and} \bibinfo{author}{\bibfnamefont{S.~I.}
  \bibnamefont{Bozhevolnyi}}, \bibinfo{journal}{Nature communications}
  \textbf{\bibinfo{volume}{5}}, \bibinfo{pages}{3809} (\bibinfo{year}{2014}).

\bibitem[{\citenamefont{Fox}(2002)}]{fox2002optical}
\bibinfo{author}{\bibfnamefont{M.}~\bibnamefont{Fox}},
  \emph{\bibinfo{title}{Optical properties of solids}} (\bibinfo{year}{2002}).

\bibitem[{\citenamefont{Tokatly and Pankratov}(1999)}]{tokatly1999hydrodynamic}
\bibinfo{author}{\bibfnamefont{I.}~\bibnamefont{Tokatly}} \bibnamefont{and}
  \bibinfo{author}{\bibfnamefont{O.}~\bibnamefont{Pankratov}},
  \bibinfo{journal}{Physical Review B} \textbf{\bibinfo{volume}{60}},
  \bibinfo{pages}{15550} (\bibinfo{year}{1999}).

\bibitem[{\citenamefont{Conti and Vignale}(1999)}]{conti1999elasticity}
\bibinfo{author}{\bibfnamefont{S.}~\bibnamefont{Conti}} \bibnamefont{and}
  \bibinfo{author}{\bibfnamefont{G.}~\bibnamefont{Vignale}},
  \bibinfo{journal}{Physical Review B} \textbf{\bibinfo{volume}{60}},
  \bibinfo{pages}{7966} (\bibinfo{year}{1999}).

\bibitem[{\citenamefont{Goldstein et~al.}(2002)\citenamefont{Goldstein, Poole,
  and Safko}}]{goldstein2002classical}
\bibinfo{author}{\bibfnamefont{H.}~\bibnamefont{Goldstein}},
  \bibinfo{author}{\bibfnamefont{C.}~\bibnamefont{Poole}}, \bibnamefont{and}
  \bibinfo{author}{\bibfnamefont{J.}~\bibnamefont{Safko}},
  \emph{\bibinfo{title}{Classical mechanics}} (\bibinfo{year}{2002}).

\bibitem[{\citenamefont{De~Ceglia et~al.}(2018)\citenamefont{De~Ceglia,
  Scalora, Vincenti, Campione, Kelley, Runnerstrom, Maria, Keeler, and
  Luk}}]{de2018viscoelastic}
\bibinfo{author}{\bibfnamefont{D.}~\bibnamefont{De~Ceglia}},
  \bibinfo{author}{\bibfnamefont{M.}~\bibnamefont{Scalora}},
  \bibinfo{author}{\bibfnamefont{M.~A.} \bibnamefont{Vincenti}},
  \bibinfo{author}{\bibfnamefont{S.}~\bibnamefont{Campione}},
  \bibinfo{author}{\bibfnamefont{K.}~\bibnamefont{Kelley}},
  \bibinfo{author}{\bibfnamefont{E.~L.} \bibnamefont{Runnerstrom}},
  \bibinfo{author}{\bibfnamefont{J.-P.} \bibnamefont{Maria}},
  \bibinfo{author}{\bibfnamefont{G.~A.} \bibnamefont{Keeler}},
  \bibnamefont{and} \bibinfo{author}{\bibfnamefont{T.~S.} \bibnamefont{Luk}},
  \bibinfo{journal}{Scientific reports} \textbf{\bibinfo{volume}{8}},
  \bibinfo{pages}{1} (\bibinfo{year}{2018}).

\bibitem[{\citenamefont{Wubs}(2015)}]{wubs2015classification}
\bibinfo{author}{\bibfnamefont{M.}~\bibnamefont{Wubs}},
  \bibinfo{journal}{Optics express} \textbf{\bibinfo{volume}{23}},
  \bibinfo{pages}{31296} (\bibinfo{year}{2015}).

\bibitem[{\citenamefont{Gradshteyn and Ryzhik}(2014)}]{gradshteyn2014table}
\bibinfo{author}{\bibfnamefont{I.~S.} \bibnamefont{Gradshteyn}}
  \bibnamefont{and} \bibinfo{author}{\bibfnamefont{I.~M.}
  \bibnamefont{Ryzhik}}, \emph{\bibinfo{title}{Table of integrals, series, and
  products}} (\bibinfo{publisher}{Academic press}, \bibinfo{year}{2014}).

\bibitem[{\citenamefont{Toscano et~al.}(2015)\citenamefont{Toscano, Straubel,
  Kwiatkowski, Rockstuhl, Evers, Xu, Mortensen, and
  Wubs}}]{toscano2015resonance}
\bibinfo{author}{\bibfnamefont{G.}~\bibnamefont{Toscano}},
  \bibinfo{author}{\bibfnamefont{J.}~\bibnamefont{Straubel}},
  \bibinfo{author}{\bibfnamefont{A.}~\bibnamefont{Kwiatkowski}},
  \bibinfo{author}{\bibfnamefont{C.}~\bibnamefont{Rockstuhl}},
  \bibinfo{author}{\bibfnamefont{F.}~\bibnamefont{Evers}},
  \bibinfo{author}{\bibfnamefont{H.}~\bibnamefont{Xu}},
  \bibinfo{author}{\bibfnamefont{N.~A.} \bibnamefont{Mortensen}},
  \bibnamefont{and} \bibinfo{author}{\bibfnamefont{M.}~\bibnamefont{Wubs}},
  \bibinfo{journal}{Nature communications} \textbf{\bibinfo{volume}{6}},
  \bibinfo{pages}{1} (\bibinfo{year}{2015}).

\bibitem[{\citenamefont{Ciraci and Della~Sala}(2016)}]{ciraci2016quantum}
\bibinfo{author}{\bibfnamefont{C.}~\bibnamefont{Ciraci}} \bibnamefont{and}
  \bibinfo{author}{\bibfnamefont{F.}~\bibnamefont{Della~Sala}},
  \bibinfo{journal}{Physical Review B} \textbf{\bibinfo{volume}{93}},
  \bibinfo{pages}{205405} (\bibinfo{year}{2016}).

\bibitem[{\citenamefont{Ding and Chan}(2018)}]{ding2018eigenvalue}
\bibinfo{author}{\bibfnamefont{K.}~\bibnamefont{Ding}} \bibnamefont{and}
  \bibinfo{author}{\bibfnamefont{C.}~\bibnamefont{Chan}},
  \bibinfo{journal}{Journal of Physics: Condensed Matter}
  \textbf{\bibinfo{volume}{30}}, \bibinfo{pages}{084007}
  (\bibinfo{year}{2018}).

\bibitem[{\citenamefont{Cirac{\`\i}}(2017)}]{ciraci2017current}
\bibinfo{author}{\bibfnamefont{C.}~\bibnamefont{Cirac{\`\i}}},
  \bibinfo{journal}{Physical Review B} \textbf{\bibinfo{volume}{95}},
  \bibinfo{pages}{245434} (\bibinfo{year}{2017}).

\bibitem[{\citenamefont{Grosso}(2003)}]{grosso2003giuseppe}
\bibinfo{author}{\bibfnamefont{G.}~\bibnamefont{Grosso}},
  \bibinfo{journal}{Solid state physics, second printing, British Library
  Cataloguing, UK}  (\bibinfo{year}{2003}).

\bibitem[{\citenamefont{Morse and Feshbach}(1954)}]{morse1954methods}
\bibinfo{author}{\bibfnamefont{P.~M.} \bibnamefont{Morse}} \bibnamefont{and}
  \bibinfo{author}{\bibfnamefont{H.}~\bibnamefont{Feshbach}},
  \bibinfo{journal}{American Journal of Physics} \textbf{\bibinfo{volume}{22}},
  \bibinfo{pages}{410} (\bibinfo{year}{1954}).

\end{thebibliography}


\end{document}